\documentclass[11pt]{article}
\usepackage{authblk}
\usepackage[utf8]{inputenc}	% Para caracteres en español
\usepackage{amsmath,amsthm,amsfonts,amssymb,amscd}
\usepackage{ulem}
\usepackage{multirow,booktabs}
\usepackage[table]{xcolor}
\usepackage{fullpage}
\usepackage{lastpage}
\usepackage{enumitem}
\usepackage{fancyhdr}
\usepackage{mathrsfs}
\usepackage{wrapfig}
\usepackage{setspace}
\usepackage{calc}
\usepackage{multicol}
\usepackage{cancel}
\usepackage{physics}
\usepackage[retainorgcmds]{IEEEtrantools}
\usepackage[margin=3cm]{geometry}
\usepackage{amsmath}

\usepackage{empheq}
\usepackage{framed}
\usepackage[most]{tcolorbox}
\usepackage{xcolor}
\colorlet{shadecolor}{orange!15}
\theoremstyle{definition}

\def\bk{{\vec K}}
\def\bk{{\vec k}}
\def\CP{{\rm CP}}

\def\bA{{\vec A}}
\def\bB{{\vec B}}

\def\ba{{\vec a}}
\def\bc{{\vec c}}

\def\bb{{\vec b}}
\def\bK{{\vec K}}
\def\bq{{\vec q}}

\def\bR{{\vec R}}
\def\bG{{\vec G}}
\def\bd{{\vec d}}
\def\bB{{\vec B}}
\def\bP{{\vec P}}

\def\bn{{\vec n}}
\def\bm{{\vec m}}
\def\bd{{\bf d}}
\def\br{{\vec r}}

\def\bsigma{{\boldsymbol \sigma}}
\def\bnabla{{\boldsymbol \nabla}}

\def \bd{{\boldsymbol d}}

\renewcommand{\vec}[1]{\boldsymbol{#1}}

\def\tr{\mathop{\mathrm{tr}}}

\def\T{\mathcal{T}}
\def\L{\mathcal{L}}
\def\D{\mathcal{D}}

\def\P{\mathcal{P}}

\def\H{\mathcal{H}}
\def\K{\mathcal{K}}

\def\diag{{\rm diag}}

\def\U{{\rm U}}

\def\SU{{\rm SU}}

\newcommand{\ov}{\overline}

\renewcommand{\Im}{\rm{Im}}

\newcommand{\beq}{\begin{equation}}
\newcommand{\eeq}{\end{equation}}
\newcommand{\beqarray}{\begin{eqnarray}}
\newcommand{\eeqarray}{\end{eqnarray}}

\newcommand{\EK}[1]{\textcolor{magenta}{#1}}

\begin{document}
\setcounter{section}{-1}
\title{\LARGE \bf  
%Notes on a Strong Coupling Theory of Magic-Angle Graphene
%Lecture Notes on Magic Angle TBG\\
%OR \\
Strong Coupling Theory of Magic-Angle Graphene: A Pedagogical Introduction}
\author{Patrick J. Ledwith, Eslam Khalaf, Ashvin Vishwanath}
\affil{Harvard University, Cambridge, MA 02138, USA.}

\maketitle
\abstract{
We give a self contained review of a recently developed strong coupling theory of magic-angle graphene. An advantage of this approach is that a single formulation can capture both the insulating and superconducting states, and with a few simplifying assumptions, can be treated analytically.  We begin by reviewing  the electronic structure  of magic angle graphene's flat bands, in a limit that exposes their peculiar band topology and geometry. We highlight how  similarities between the flat bands and the lowest Landau level give insight into the effect of interactions. For example, at certain fractional fillings, we note the promise for realizing fractional Chern states. At integer fillings, this approach points to  flavor 
ordered insulators, which can be captured by a sigma-model in its ordered phase. Unexpectedly, topological textures of the sigma model carry electric charge which allows us to extend the same theory to describe the doped phases away from integer filling. We show how this approach can lead to superconductivity through the proliferation of charged topological textures, and estimate the T$_c$ for the superconductor. We highlight the important role played by an effective super-exchange coupling both in pairing and in setting the effective mass of Cooper pairs.  Seeking to enhance this coupling helps predict new superconducting platforms, including the recently discovered alternating twist trilayer platform. We also contrast our proposal from strong coupling theories for other superconductors. 

%Here we give a fully analytic discussion of the essential physics of magic angle graphene. While this will necessarily involve some approximations, we hope the advantage of insight provided by these simplified models will outweigh the necessarily simplified picture they will paint. Where appropriate, we will comment on more detailed calculations and how they corroborate or modify the simple picture. :
}
\thispagestyle{empty}

%\begin{center}
%{\LARGE \bf Lecture Notes on Magic Angle TBG}\\
%\end{center}

\section{Introduction}
%\PL{Collected refs that we were asked to cite via email: FQHE tight binding: \cite{Andrews2020} (cited this nearby but separate from the other FCI numerical studies).  Lattice relaxation \cite{Lucignano2019,Cantele2020} (cited this below with Koshino, Carr). Trugman Kivelson \cite{TrugmanKivelson} (cited this near pseudopotential argument)}

In early 2018, the experimental discovery of a host of novel phenomena in twisted bilayer graphene (TBG) took the physics world by storm \cite{Pablo1,Pablo2}.   New insulating states induced by electron-electron interactions, as well as robust  superconducting states were discovered.  These experiments built on  earlier work \cite{dosSantos2007,Li2010,Laissardiere2010,Mele2010,Lucian2011,Bistritzer2011,dosSantos2012,Wong2015,Kim2017,Huang2018,Rickhaus2018} that pointed to the existence of exceptionally narrow and isolated bands arising from the moire pattern obtained on twisting the two graphene sheets close to a  `magic angle', i.e. magic angle TBG (MATBG). The condensed matter community had perhaps not witnessed such a tectonic shift since the experimental discovery of high temperature superconductivity, more than three decades earlier. 

There were  other parallels between the two discoveries. Although  MATBG with a superconducting transition temperature of a few Kelvin may seem like an unlikely candidate for the title of a high T$_c$  superconductor,  a proper comparison of scales is required. At the magic angle of ~1.1 degree ($\theta \sim 1/50$th of a radian), many relevant scales  - from the lattice spacing to the size of the Coulomb interaction - are scaled by this small angle. Thus, a better comparison with the cuprates or other correlated solids,  is  obtained by  scaling $T_c$ by the  small angle (expressed in radian)  $T_c/\theta$, which yields values consistent with  the high $T_c$ family. Additionally, in both cases, interacting insulating states and superconductors appear in conjunction in typical phase diagrams and strange metal transport has been reported on raising temperature (see eg. ref \cite{Balents2020} for a review).

At the same time,  there are several important distinctions.  A principle distinction that will be emphasized here is the nontrivial quantum band geometry of the flat bands of MATBG. This is a special feature arising from graphene's Dirac electrons. On the experimental side, the most striking manifestation of this feature is the emergence of spontaneous integer quantum Hall states (anomalous quantum Hall) under certain conditions in twisted bilayer systems. This remarkable unification of quantum Hall and high Tc superconductivity in a single experimental platform, points to this underlying unique feature of TBG.

In  1987,  shortly after the discovery of the cuprate high temperature superconductors, Anderson \cite{ANDERSON1196} published a short paper    ``The Resonating Valence Bond (RVB) Theory of Superconductivity" where he outlined the foundations of a theoretical program for the cuprates. Along with several other key ideas, it was proposed that (i) the underlying model was the Hubbard Model with hopping $t$ and onsite Coulomb repulsion $U$ that would give rise to local magnetic moments with an antiferromagnetic  coupling  $J\sim t^2/U$ (super-exchange) between them and (ii) the seeds of superconductivity were already contained in the insulator and charge doping simply released  the singlets already created by the superexchange interaction.These singlets then condensed into the superconducting state. Consequently, at low doping the superconducting T$_c$ was predicted to be limited by phase fluctuations and not the pairing gap, as in BCS superconductors.  Notwithstanding the fact that later experiments have painted a rich and complex picture of the cuprates, these seminal ideas made a deep impression that lasts to this day. 
%Anderson The Resonating Valence Bond State in La2CuO4 and Superconductivity \cite{ANDERSON1196}

A direct attempt to translate this program to MATBG immediately runs into problems. As was noticed starting very early on \cite{Po2018, Po2018faithful, Ahnetal, Songetal}, writing down a tight binding model that captures the flat bands alone, while preserving all relevant symmetries, runs into a topological obstruction. One must ether give up on some symmetries, such as the 180 degree rotation symmetry $C_{2z}$, time reversal $\mathcal T$,  or conservation of valley charge (valley U(1)) \cite{KangVafek2018} or augment the model with additional bands \cite{Carr, Po2018,Po2018faithful}. Note, preserving these symmetries is not simply a technical nicety - they are the precise symmetries responsible for the band touching of the two bands of magic angle graphene. What then plays the role of the underlying physical model, analogous to the Hubbard model employed for the cuprates? 

In the first section of this review we outline an approach to studying the single particle eigenstates directly in momentum space and advocating for a choice of basis that will make the subsequent discussion, on including  interaction effects, natural. This is made explicit by tuning to a particular `chiral limit', where the model has remarkably simple features. In fact on tuning angle, perfectly flat bands are obtained, and even the form of the single particle wavefunctions in this limit can be derived (almost entirely) from analytic arguments \cite{Tarnopolsky}. A close analogy to quantum Hall wavefunctions is pointed out; this model has several attractive mathematical features that remain to be fully understood \cite{becker2020, Popov2020,wang2020chiral,naumis2021reduction,Ren_2021,sheffer2021}. The central observation here is that one can make linear combination of bands that are simultaneously both sublattice polarized and carry unit Chern number. This property holds even away from the chiral limit and forms the starting point for a strong coupling theory  of integer filling in Section 2.  
%Within the Hubbard model analogy,  the chiral model is like the Hu only nearest neighbor hopping "t",while additional hopping most notably second neighbor  t' changes the properties and can be imprtant.) 

The strong coupling theory, when applied to integer fillings, predicts insulating states which can be thought of as generalized flavor ferromagnets. In all, there are {\em eight} nearly flat bands. A fourfold degeneracy is attributed to spin and valley. In addition a further twofold band degeneracy is present, corresponding roughly to the two sublattices of graphene. We show how in an idealized limit a flavor ferromagnet is predicted, and a large emergent symmetry relates various flavor orders. On moving away from the ideal limit, various anisotropies  emerge that select a subset of flavor ordered ground states. The ground state is thus a relatively conventional state, well approximated by Hartree Fock wavefunctions, and is to be  contrasted with the exotic insulator proposed by Anderson for the cuprates, an RVB quantum spin liquid \cite{ANDERSON1196}. However, in one respect they do fit within the Anderson viewpoint. Somewhat to our surprise \cite{SkPaper}, it was found that doping the insulators did not require adding any additional charge degrees of freedom. Instead, charge excitations in the form of skyrmions were already present in the model. Further, it was realized that in certain cases these were nothing but Cooper pairs which can condense, leading to an all electronic mechanism for superconductivity.  The transition temperature for this mechanism is calculated in a simplified limit and the results are broadly in agreement with experimental data in the low doping limit. Numerical DMRG simulations that capture essential features of this model reveal that indeed the presence of low energy skyrmion excitations near integer filling is closely tied to the appearance of superconductivity at finite doping \cite{Chatterjee}.  This is the subject of Section 3. 
%\AV{We also discuss supporting numerics? DMRG?} 
In addition to this detailed mechanism, a more basic similarity is the presence of a superexchange interaction. We will see that we can understand the eight flat bands of MATBG as four bands with Chern number $+1$ and four bands with Chern number $-1$. The band dispersion acts as a virtual tunneling between opposite Chern sectors and therefore generates a superexchange interaction that favors states that are antiferromagnetic in Chern sector.
%energy scales argument relates the pairing to the underlying microscopic energy scales - discuss $J$.... 

Let us further contrast the mechanism outlined above with the RVB idea,  by placing both in a broader context. We note that in the modern parlance the RVB state has topological order and fractionalization in the precise sense that it has emergent excitations with unusual statistics and quantum numbers. This obviously goes beyond the idea of classical Landau order parameters. In contrast the insulators we propose for TBG are flavor ordered phases, without fractionalization. However, thanks to the nontrivial bands of magic angle graphene, these orders are not entirely classical - for instance topological textures such as skyrmions carry electric charge. Such spontaneous symmetry breaking in topological bands has previously been discussed in relation to superconductivity \cite{abanovwiegmann, GroverSenthil}, however a microscopic realization of the relevant type of insulator itself has proved elusive. Here we explain how the flavor ordered states of MATBG at even integer filling fulfill these requirements. 

Naturally, predictions for new platforms for superconductivity should emerge from a deeper understanding of the physical mechanism. For cuprate superconductors, once key ingredients were identified, a class of  nickelate materials were deemed promising analogs and have recently been fabricated and shown to exhibit superconductivity \cite{Li2019}. In the strong coupling approach to MATBG described here, the principle goal is to enhance a superexchange coupling $J$, which in the typical settings of graphene Moir\'e systems is related to $C_2$ symmetric structures. While there is already a vast library of graphene based Moir\'e materials, MATBG is one of the few that exhibits this symmetry. Alignment with a substrate of hBN for example leads to symmetry lowering. A rare class of  structures which have all the requisite ingredients for the strong coupling mechanism described above  are the alternating twist multilayers \cite{Khalaf_multilayer}. This is the subject of Section 4, including both trilayer and tetralayer, the first of which was recently reported in experiment to possess robust superconductivity \cite{Hao2021TTG,Park2021TTG}.

%\begin{shaded}
%\subsection{Outline}
%\begin{enumerate}
%\item {\bf PART 1} Introduce a model for the band structure of magic angle graphene which can be analytically solved. Highlight the similarities with Landau level wavefunctions. 

%\item {\bf PART 2} Introduce interactions and the soluble limit where U(4)x U(4) symmetry emerges and ground state can be found. Discuss simplification to spinless case U(2)xU(2) case and  anisotropy calculations. Discuss derivation of the sigma models and calculation of topological term.  
%\begin{equation}
%\begin{split} 
%    {\mathcal  L}[\hat {n}_+,\,\hat {n}_-] = \sum_{i=\pm} {\mathcal A}[\hat {n}_i] \cdot \frac {d \hat{n}_i}{dt} -\frac{\rho}{2} |\nabla{\hat n}_i|^2\\
%    -J {\hat n}_+\cdot {\hat n}_- +(J+\lambda) n^z_+n^z_- \nonumber
%    \end{split}
%    \end{equation}
%\item {\bf PART 3} Discuss charged skyrmions, their energy versus the particle-hole excitation energy. Show that in the disordered phase of the sigma model superconductivity emerges.
%\end{enumerate}
%\end{shaded}

\section{Single Particle Electronic Structure}
%\AV{A few line summary of the most important points in this section?eg. Dirac electrons see a Moire pattern which impacts their tunneling between graphene sheets in a periodic fashion. this leads to band structures with unusual momentum space geometry and topology - setting the stage for interaction effects.  }
\label{Sec:SingleParticle}
\begin{shaded}
{In this section we describe how to go from two decoupled graphene sheets to the topological flat bands of twisted bilayer graphene. We start from the low energy graphene Dirac cones in each layer and couple them through tunneling between layers. This tunneling has a moir\'{e} periodicity that is reflected in the twisted bilayer graphene band structure. Finally, we describe a ``chiral" limit where tunneling at AA stacking sites is neglected. In this limit we find that the wavefunctions are very similar to those of the lowest Landau level (LLL). This structure of the wavefunctions has implications for the possibility of fractional Chern insulating (FCI) states in twisted bilayer graphene. It also sets the stage for treating interactions at integer filling.}
\end{shaded} 

To set notation, let us begin with the electronic structure of graphene. It consists of two bands that cross at two Dirac cones at the $K$ and $K'$ points in the Brillouin zone (BZ). There is one electron per unit cell, and so accounting for spin we should fill up the bottom band. The low energy dispersion thus consists of the Dirac cones

% The electron dispersion at charge neutrality (the charge neutrality point, where one electron per atom is present, however due to spin this corresponds to filling half the states in the Brillouin Zone). The dispersion can be approximated by a Dirac equation with the following dispersion. 

\begin{equation}
    H = v_F \left ( p_x\, \sigma_x \otimes \tau^z +p_y \, \sigma_y \otimes \tau^0\right ) 
\end{equation}

where $\sigma_z =\pm 1$ corresponds to the A, B sublattices of graphene and $\tau_z =\pm 1$ corresponds to the $K,\,K'$ valleys of graphene.
% \begin{note}
% \textbf{Include a discussion of graphene's symmetries and implementation on the wavefunctions. Set up two site unit cell, real space lattice and reciprocal lattices.}.
% \end{note}

%\subsection{Review: Single Particle Band Structure}
Now consider the twisted bilayer in the continuum limit, focusing on a single valley; {we will derive the Bistritzer-Macdonald (BM) Hamiltonian \cite{Bistritzer2011}}. We can write the Hamiltonian in the two layers as:
\begin{equation}
     H_+ = \begin{bmatrix}
H_{UU} & H_{UD}\\
H_{DU} & H_{DD}
\end{bmatrix}
\end{equation}

We begin with independent layers and then couple them together. The Up and Down layer Dirac Hamiltonians are:
\begin{equation}
\begin{split}
H_{UU} = v_F ({-i \bf \bnabla-K^U})\cdot {\bf \bsigma}_{\theta/2}\\
H_{DD} = v_F ( {-i\bf \bnabla-K^D})\cdot {\bf \bsigma}_{-\theta/2}
\end{split}
\label{HUUDD}
\end{equation}
where we have taken into account the shift of the Dirac points and the Pauli matrices due to rotation so that ${\bf \sigma}_{\theta/2} =e^{-i\frac\theta4\sigma_z} (\sigma_x,\,\sigma_y) e^{+i\frac\theta4\sigma_z}  $. Note that we really want to rotate $\bnabla - \bK^{U,D}$ but it is equivalent and more convenient to rotate $\bsigma$ instead. We now couple the Dirac points together by interlayer hopping terms $H_{DU} = H^\dagger_{UD}$ where

\begin{equation}
\begin{split}
H_{UD} = T_0(\br)\sigma^0 +T_{AB}(\br) \sigma^+ + T_{BA}(\br) \sigma^-
\end{split}.
\end{equation}

Let us now derive the form of the tunneling matrix elements - for better intuition first consider very local tunneling (i.e. electrons tunnel only when the atoms of the two layers occlude each other). We then need to consider three cases, where (i) the atoms of both sublattices are on top of each other, the so called AA or equivalently the BB regions which form a triangular lattice, see Fig. 1; (ii) where the A atoms of the lower layer align with the B atoms of the upper layer (AB regions), they form one sublattice of a honeycomb lattice on the moire scale; and finally (iii) the  BA regions, where B atoms of the lower layer align with the A atoms of the upper layer and form the other sublattice of the honeycomb lattice on the moire scale.

Tunneling, say in the $AA$ regions, could be represented as:
$T_0 \sim \sum_{n}\delta(\bf {r-R}_n)$ where the sum runs over the triangular Moir\'e lattice sites locations. From the Poisson resummation formula this can be equivalently written as $\sum_{n}\delta(\bf {r-R}_n) = \frac{1}{A_M}\sum_m e^{-i{\bf G_m \cdot r}}$, where $A_M$ is the area of the moire unit cell and $\bf G_m$ are the moire lattice reciprocal vectors. On the other hand the tunneling matrix elements in the other regions acquire additional phase factors due to their displacement from the unit cell centers. They are $T_{AB} \sim \sum_{n}\delta({\bf r-R_n - r}_A) = \frac{1}{A_M}\sum_m e^{i{\bf G_m \cdot r}_A} e^{-i{\bf G_m \cdot r}}$ and $T_{BA} \sim \sum_{n}\delta({\bf r-R_n - r}_B) = \frac{1}{A_M}\sum_m e^{i{\bf G_m \cdot r}_B} e^{-i{\bf G_m \cdot r}}$. In reality, one has to take into account a more general interlayer hopping $T(\bf r)$ with Fourier components $\tilde{T}({\bf G_m})$. Since the interlayer hopping  depends on the 3D separation between atoms, and the separation between layers is more than twice the intra-layer atomic separations, we expect the Fourier components to decrease rapidly and we can keep the lowest order harmonics \cite{BM}. 
% {\color{red} I believe this is due to the fact that away from eg. AA sites the atoms are not in registry but randomly located within an atomic seperation, but the distance between the layers means this contrast is not strongly reflected in the tunneling. Hence the lowest harmonic approximation - makes sense?} 
To obtain the tunneling strengths for the lowest order harmonics we will appeal to rotation symmetry. Note the original choice of $K^{U/D}$ for the Dirac points breaks rotation symmetry. To make rotation symmetry manifest, we will move the Dirac points in both layers to the origin using the unitary transformation with diagonal entries: $W = {\rm diag} ( e^{i {\bf K^U\cdot r}},\, e^{i{\bf K^D\cdot r}})$. This leads to the following transformed Hamiltonian:
\begin{equation}
\begin{split}
H'_{UU} = v_F (-i \bnabla)\cdot {\bsigma}_{\theta/2},\\
H'_{DD} = v_F ( -i\bnabla)\cdot \bsigma_{-\theta/2}.
\end{split}
\end{equation}

We define $\bq_1 = \bK^U-\bK^D = k_\theta(0,-1)$, and the vectors related by $2\pi/3$ rotations $\bq_{2,3} = k_\theta(\pm \sqrt{3}/2, 1/2)$. Then we can write the tunneling matrix elements as
\begin{equation}
\begin{aligned}
H'_{UD} & = w_0 U_0(r)\sigma^0 +w_1 U_{AB}(r) \sigma^+ + w_1 U_{BA}(r) \sigma^-,\\
U_0 (r) & = e^{-i{\bf q_1\cdot r}}+e^{-i{\bf q_2\cdot r}}+e^{-i{\bf q_3\cdot r}},\\
U_{AB} (r) & = e^{-i{\bf q_1\cdot r}} + e^{i\frac{2\pi}{3}}e^{-i{\bf q_2\cdot r}}+ e^{i\frac{4\pi}{3}} e^{-i{\bf q_3\cdot r}},\\
U_{BA} (r)& = e^{-i{\bf q_1\cdot r}} + e^{-i\frac{2\pi}{3}}e^{-i{\bf q_2\cdot r}}+ e^{-i\frac{4\pi}{3}} e^{-i{\bf q_3\cdot r}}.
\end{aligned}
\end{equation}
% {\bf estimate for w?}

The BM Hamiltonian has the translation symmetry 
\begin{equation}
H(\br + \ba_{1,2}) = V H(\br) V^\dag  \qquad V = \diag(1,e^{-i \phi}, 1, e^{-i \phi}),
\label{translations}
\end{equation}
where the moir\'{e} lattice vectors $\ba_{1,2}$ and reciprocal lattice vectors $\bb_{1,2}$ are
\begin{equation}
\ba_{1,2} = \frac{4\pi}{3k_\theta}\left( \pm \frac{\sqrt{3}}{2}, \frac{1}{2} \right), \qquad
\bb_{1,2} = \bq_{2,3} - \bq_1 = k_\theta\left(\pm\frac{\sqrt{3}}{2}, \frac{3}{2} \right). 
\end{equation}
We will find it useful to sometimes use triangular lattice and reciprocal lattice coordinates
\begin{equation}
    r_i = \frac{\br \cdot \bb_i}{2\pi} \qquad k_i = \frac{\bk \cdot \ba_i}{2\pi}
\end{equation}
which satisfy $\br = r_1 \ba_1 + r_2 \ba_2$ and $\bk = k_1 \bb_1  + k_2 \bb_2$.

The relative phase shift between the top and bottom layers under translations in \eqref{translations} can be understood from the misalignment of their graphene Brillouin zones due to the twist angle $\theta$.  In particular, momenta in the bottom layer are shifted by $\bq_1$ relative to the top layer which generates the phase shift $e^{i\bq_1 \cdot \ba_{1,2}} = e^{-i\phi}$ under translations. This is encoded in the form of the Bloch states
\begin{equation}
    \psi_{\bk}(\br) = \begin{pmatrix} u^U_{\bk}(\br) e^{i(\bk - \bK)\cdot \br}  \\ u^D_{\bk}(\br) e^{i(\bk - \bK')\cdot \br}\end{pmatrix}_{\rm Layer},
    \label{blochstates}
\end{equation}
where $\bK' = \bK  - \bq_1$ is the Moir\'{e} $K'$ point wavevector in terms of the Moir\'{e} $K$ point wavevector.  The functions $u^{U,D}_{\bk}(\br)$ are periodic under $\br \to \br + \ba_{1,2}$. For zero tunneling, or large angles, the states at $K$ ($K'$) correspond to the top layer and bottom graphene layer respectively.

Note, in the simplest approximation $w_0=w_1 \sim 110$  meV. However, later work \cite{Koshino, Carr,Lucignano2019,Cantele2020} showed that out of plane lattice relaxation, followed by in plane relaxation, reduce the ratio $\kappa = w_0/w_1$ from unity to $\kappa=0.8$ and then further to $\kappa = 0.65 - 0.75$. This quantity will play an important role and we will often treat it as a free parameter.

\subsection{The Chiral Model: A simple limiting case}

{It will be particularly interesting to consider the {\bf chiral limit}, where $\kappa=0$. As we show in this section, in this limit, there exists an exactly flat zero energy flat band \cite{Tarnopolsky}. Furthermore, the wavefunctions are very similar to those of the LLL; they have identical quantum geometry in a sense that we will make precise. This result implies that we can expect chiral twisted bilayer graphene to host fractional Chern insulator ground states, at least for a sufficiently short range interaction potential \cite{Ledwith}.

We begin with the Bistritzer-Macdonald (BM) model discussed above
\begin{equation}
    H = \begin{pmatrix} -i v \bsigma_{\theta/2} \cdot \bnabla & T(\br ) \\
    T^\dag (\br) & -iv \bsigma_{-\theta/2} \cdot \bnabla \end{pmatrix}
    \label{ham}
\end{equation}
with 
\begin{equation}
    \begin{aligned}
        T(\br) & = \begin{pmatrix} w_0 U_0(\br) & w_1 U_1(\br) \\ w_1 U^*_1(-\br) & w_0 U_0(\br) \end{pmatrix}, \\
        U_0(\br) & = e^{-i \bq_1 \cdot \br } + e^{-i \bq_2 \cdot \br } + e^{-i \bq_3 \cdot \br },  \\
        U_1(\br) & = e^{-i \bq_1 \cdot \br } + e^{i \phi}e^{-i \bq_2 \cdot \br } + e^{-i \phi}e^{-i \bq_3 \cdot \br },
        \label{tunnel}
    \end{aligned}
\end{equation}
The vectors $\bq_i$ are $\bq_1 = k_\theta(0,-1)$ and $\bq_{2,3} =k_\theta(\pm \sqrt{3}/2,1/2)$ and $\phi = 2\pi/3$. 

For the following discussion we choose units where $v = k_\theta = 1$. We will also find it useful to choose the origin of the Moir\'{e} BZ to be the $K$ point to simplify some later expressions, even though the most symmetric choice is the $\Gamma$ point.

A very useful approximation to the above Hamiltonian is to set $\kappa = 0$, such that tunneling only occurs in AB regions. Realistically we expect $\kappa \approx 0.7 < 1$ due to relaxation effects.  While decreasing it all the way to zero is a drastic approximation, we will find it useful to do so at first and then examine the effects of going back to a realistic value of $\kappa$ later in these notes.  The model with $\kappa = 0$ is known as the chiral model because it has the chiral
symmetry $\sigma_z  H  \sigma_z = -H $. That is, given an eigenstate $|\Psi\rangle $ of $H$ with energy $E$, one can generate an eigensate $\sigma_z |\Psi\rangle$, an eigenstate with energy $-E$. Further, zero energy eigenstates can be chosen to be eigenstates of $\sigma_z$, i.e. they are sublattice polarized. Below, we will show the existence of special (magic) angles where the entire band is at zero energy. As a consequence we can define a sublattice polarized basis for these eigenstates. What is less obvious is that these sublattice polarized bands each have a Chern number $\pm1$ \cite{Tarnopolsky,KIVCpaper}, and in fact share many similarities, in ways that will be made precise below, with the LLL \cite{Ledwith}.  

The chiral symmetry is especially useful at the magic angle; here we obtain \emph{exactly} flat bands at zero energy that are eigenstates of $\sigma_z$. Since we are interested in zero modes that are eigenstates of $\sigma_z$, we rewrite the Hamiltonian by exchanging the second and third columns after which $\sigma_z = \diag(1,1,-1,-1)$ and 
\begin{equation}
    H = \begin{pmatrix}0 & D^\dag \\ D & 0 \end{pmatrix} \quad D = \begin{pmatrix} -2i \ov \partial & \alpha U_1(\br) \\ \alpha U_1(-\br) & -2i \ov \partial \end{pmatrix},
    \label{chiral ham}
\end{equation}
with $\alpha = w_1/vk_\theta$. Because $\{ H ,\sigma_z \} = 0$, zero energy solutions may be chosen to be eigenstates of $\sigma_z$. We therefore search for solutions to the equation $D \psi_\bk(\br) = 0$ for each Moir\'{e} Bloch momentum $\bk$. The solutions when $\sigma_z = -1$ can be obtained using $C_2 \T$ symmetry: $\chi_\bk(\br) = \psi_{\bk}^*(-\br)$. 

\subsubsection{Zero Modes of Chiral Model:} When $\bk = K$, we are guaranteed a zero mode solution by continuity from $\alpha = 0$, i.e. the unperturbed Dirac points in the top layer. 
This is because the $\alpha = 0$ zero mode $\psi_K = \begin{pmatrix} 1 & 0 \end{pmatrix}^T$ is pinned to zero energy: it is an $+1$ eigenstate of $\sigma^z$ and it cannot mix with the $-1$ eigenstate because they have different eigenvalues of $C_3 = e^{i \phi \sigma_z}$. It also cannot move to different values of $\bk$ because it is pinned to $\bk
= K$ by $C_3$ symmetry; there are precisely three $C_3$ symmetric points in the mBZ ($K, K', \Gamma$) and the zero mode cannot jump discontinuously from one point to another while $\alpha$ varies continuously. Thus $\psi_K$ has zero energy for all $\alpha > 0$.

Next we note that because $D$ only has antiholomorphic derivatives 
\begin{equation}
    D\left(F(z) \psi_K(\br)\right) = F(z) D \psi_K(\br) = 0
\end{equation}
for any holomorphic function $F(z)$ where $z = x+iy$. However, any holomorphic function $F(z)$ is either constant or blows up at infinity by Liouville's theorem. Hence, this state cannot be a Bloch state at a momentum other than $K$. At this point it may appear that we cannot obtain zero modes except at isolated crystal momenta. However there is way out -  we can allow $F(z)$ to be a meromorphic function. Although such a function would necessarily have poles at certain positions in the unit cell, we could arrange for them to be precisely cancelled by zeros in $\psi_K(\br)$. Note, since this is a spinor wavefunction, we will need both components of $\psi_K(\br)$ to simultaneously vanish at that location in the unit cell. We can explore if such a condition is satisfied by varying the angle. Indeed the magic angles are
    precisely those angles for which $\psi_K$ has a zero in both of its components at some $\br$. 
    
    \begin{figure}
        \centering
        \includegraphics[width = 0.5\textwidth]{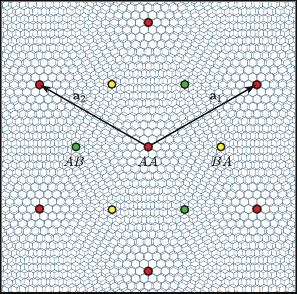}
        \caption{Real space visualization of the moir\'{e} pattern and depiction of the AA stacking $\br = 0$, AB stacking $\br = -\br_0$, and BA stacking $\br = \br_0$ points.  All three of these points map to themselves up to shifts by lattice vectors under $C_3$ symmetry}
        \label{fig:realspace}
    \end{figure}
    
    In fact, symmetry helps us locate where the zero will appear.  We focus on the points $\pm \br_0$, the BA/AB stacking points respectively with $\br_0 = \frac{1}{3}(\ba_1 - \ba_2)$. These points are distinguished by $C_3$ symmetry since they map to themselves up to a lattice vector, $C_3 \br_0 = \br_0 + \ba_2$, see Fig.\ref{fig:realspace}. We show that three fold rotation symmetry implies that one of the components $\psi_{K2}(\pm \br_0) = 0$.   Indeed, $C_3$ symmetry implies $\psi_K(R_\phi \br) = \psi_K(\br)$ where $R_\phi$ rotates vectors by an angle $\phi$. One may deduce that this is the correct representation by continuity from $\alpha = 0$ where $\psi_K = (1, 0)^T$. However, we additionally have 
    \begin{equation}
        \psi_{K2}(\pm \br_0) = \psi_{K2}(\pm R_\phi \br_0) = \psi_{K2}(\pm\br_0 \pm \ba_2) = e^{\mp i\phi}\psi_{K2}(\pm \br_0)
    \end{equation}
    since due to the form of the Bloch states \eqref{translations} the second component of $\psi_K$ picks up a phase under translations.  
    
    While it is possible to show that $D\psi_K = 0$ does not allow for $\psi_{K1}(-\br_0) = 0$, it is possible for the single component  $\psi_{K1}(\br_0)$, which can always be chosen real, to vanish for certain magic angles. Thus we have one tuning parameter to set a single real number to zero, which generically will give a series of points as solutions, i.e. magic angles. The first of these angles is at nearly the same angle as realistic TBG and can be approximated using perturbation
    theory for small $\alpha$.  In particular, we plug in the ansatz
    \begin{equation}
        \psi_K = \begin{pmatrix} 1 + \alpha^2 u_2 + \cdots \\ \alpha u_1 + \cdots \end{pmatrix}
    \end{equation}
    to the zero mode equation $D \psi_K = 0$ and obtain
    \begin{equation}
        u_1 = -i( e^{i \bq_1 \cdot \br}+ e^{i \bq_2 \cdot \br} + e^{i \bq_3 \cdot \br}   ) \quad
        u_2 = \frac{i}{\sqrt{3}} e^{-i \phi}(e^{-i \bb_1 \cdot \br} + e^{i \bb_2 \cdot \br}+  e^{i (\bb_1 - \bb_2) \cdot \br}) + \rm{c.c.}
    \end{equation}
    We then compute $u_1(\br_0) = 0$ and $u_2(\br_0) = -3$ such that when $\alpha \approx 1/\sqrt{3} \approx 0.577$ the entire spinor vanishes.  The perturbation expansion may be carried out to higher orders and seems to get arbitrarily close to the numerically obtained value of $\alpha_1 \approx 0.586$ for the first chiral magic angle.  There seem to be infinitely many $\alpha$ for which $\psi_K$ vanishes with the quasiperiodicity $\alpha_n \approx \alpha_{n-1} + 1.5$ holding for large
    $\alpha$. Other metrics such as vanishing of Dirac velocity, maximum gap to neighboring bands and minimum bandwidth also gives the same result. See Ref. \cite{Tarnopolsky} for more details. 

    \subsubsection{Wavefunctions and topology:} To generate the rest of the states in the band we need to choose a meromorphic function $F_\bk(z)$ that is Bloch periodic. To do so, we write $F_\bk(z) = f_\bk(z)/g(z)$ where both $f_\bk$ and $g$ are holomorphic and $g(z_0 + m a_1 + n a_2) = 0$ where $z_0 = r_{0x} +i r_{0y}$ and similarly $a_{1,2}$ are the complex number versions of the lattice vectors $\ba_{1,2}$. The only choice for $g(z)$, up to multiplication by an exponential function, is the Jacobi theta function.
    \begin{equation}
        g(z) = \vartheta_1\left(\frac{z-z_0}{a_1} \bigg|\omega \right),  \qquad \vartheta(u | \tau ) = -i\sum_{n = - \infty}^\infty (-1)^n e^{\pi i \tau \left(n+ \frac{1}{2} \right)^2 + \pi i(2n+1)u },
        \label{denom}
    \end{equation}
where $\omega = e^{i \phi}$. Under translations the Jacobi theta function satisfies
\begin{equation}
\vartheta_1(u + 1|\tau) = - \vartheta_1(u|\tau), \qquad
\vartheta_1(u + \tau|\tau) = -e^{-\pi i\tau - 2\pi i u} \vartheta_1(u|\tau).
\label{transprops}
\end{equation}
This quasiperiodicity under translations is the main reason the Jacobi theta function appears here; it ensures the zeros are in the same place for each unit cell and it will be crucial for reproducing Bloch periodicity. Indeed, a translation by $\ba_1$ shifts argument of the theta function by $1$ while a translation by $\ba_2$ shifts the argument of the theta function by $a_2/a_1 = e^{i\phi} = \tau$). Note that $\vartheta_1(0|\tau) = 0$ such that $g(z)$ has the desired zeros. The properties \eqref{transprops} enable us to construct Bloch states by choosing $f_\bk(z)$ to also be a theta function, since although a single theta function is only an eigenstate under translations in one direction, the above $u$ dependence in translations proportional to $\tau$ will always cancel out in any ratio of theta functions leaving a pure phase. This pure phase may be manipulated by shifting the arguments of the theta functions relative to each other, and an exponential function can also be included as an additional knob. We may therefore engineer $f_{\bk}(z) = e^{2\pi i k_1 z/a_1}\vartheta_1((z-z_0)/a_1 - k/b_2 | \omega)$ for $k = k_x + i k_y$ (measured from the $K$ point) such that
\begin{equation}
    \begin{aligned}
    \psi_{\bk}(\br) & = e^{2\pi i k_1 z/a_1} \vartheta_1\left(\frac{z-z_0}{a_1} - \frac{k}{b_2} \bigg| \omega \right)\frac{\psi_{K}(\br)}{\vartheta_1\left(\frac{z-z_0}{a_1} | \omega \right) } , \\
    u_\bk(\br) & = e^{-2\pi i r_2 k /b_2} \vartheta_1\left(\frac{z-z_0}{a_1} - \frac{k}{b_2} \bigg| \omega \right)\frac{u_{K}(\br)}{\vartheta_1\left(\frac{z-z_0}{a_1} | \omega \right) },
    \label{chtbg_wfs}
\end{aligned}
\end{equation}
are Bloch periodic and periodic under translations respectively. Note that we take $u = \begin{pmatrix} u^U & u^D \end{pmatrix}^T$; see \eqref{blochstates}.

Let us make two observations - first, note the functional dependence of Bloch wavefunctions $u_\bk(\br)$ on moving through the BZ. In the gauge we have chosen, the entire dependence is via $k =k_x+ik_y$, i.e. it is an {\em analytic} function of $k$. Next, note that these wavefunctions describe a Chern band with $C = +1$. To see this, we note that $u_{\bk}(\br)$ is well defined for all $\bk$ and $\br$ with no singularities, but it is not periodic under $\bk \mapsto \bk+\bb_{1,2}$.  Instead, using \eqref{transprops} we have 
\begin{equation}
\begin{aligned}
    u_{\bk+\bb_{1}}(\br) & = -e^{-2\pi i r_1}e^{i\pi \omega + 2\pi i z_0} e^{2\pi i k/b_2} u_\bk(\br), \\
    u_{\bk+\bb_{2}}(\br) & = -e^{-2\pi i r_2} u_\bk(\br)
    \label{kspace_trans}
\end{aligned}
\end{equation}
For the Berry connection $\bA(\bk) = -i\bra{\tilde{u}_\bk}\bnabla_{\bk} \ket{\tilde{u}_\bk}$, where $\tilde{u} = u/\norm{u}$ is a normalized wavefunction, this implies 
\begin{equation}
    \bA(\bk + \bb_1) = \bA(\bk) + \frac{1}{2} \ba_1 + \ba_2, \qquad \bA(\bk + \bb_2) = \bA(\bk) 
    \label{berryconn_periodicity}
    \end{equation}
    For both \eqref{kspace_trans} and \eqref{berryconn_periodicity} we used $b_1/b_2 = -\omega$, and for \eqref{berryconn_periodicity} we used triangular lattice coordinates where $2\pi \bnabla_\bk = \ba_1 \partial_{k_1} + \ba_2 \partial_{k_2}$. The Chern number can be evaluated as
    \begin{equation}
    \begin{aligned}
        2\pi C & = \int_{\rm BZ} \bnabla \times \bA = \oint_{\partial \rm BZ} \bA \cdot d \bk \\
        & = \int_0^1 \left( \bA(k_1 \bb_1) - \bA(k_1 \bb_1 + \bb_2) \right)\cdot \bb_1 dk_1
          + \int_0^1 \left(\bA(\bb_1 + k_2\bb_2) - \bA(k_2 \bb_2) \right)\cdot \bb_2 dk_2 \\
        & = \ba_2 \cdot \bb_2 = 2\pi,
        \end{aligned}
    \end{equation}
and so $C = 1$. The $C_2 \T$ related band with $\sigma_z = -1$ then has $C = -1$.  Including the bands from the other graphene valley and spin degeneracy, we obtain four exactly flat bands with $C = +1$ and four exactly flat bands with $C = -1$.
The reader may question how meaningful it is to assign energetically degenerate bands equal and opposite Chern numbers. Mathematically, there is of course no issue with this statement since the Chern number only depends on the band's wavefunctions and not its energetics. Physically, we may imagine splitting the bands; this splitting may come explicitly from a sublattice potential (induced by hBN substrate alignment) but in general may also come from interaction induced spontaneous symmetry breaking. Note also that the opposite Chern sectors are related by $C_2 \T$ symmetry. Furthermore, we will see in later parts of the review that thinking in terms of degenerate Chern bands can be very useful for understanding interacting physics even if the many body states have net Chern number zero.

\begin{figure}
    \centering
    \includegraphics[width = \textwidth]{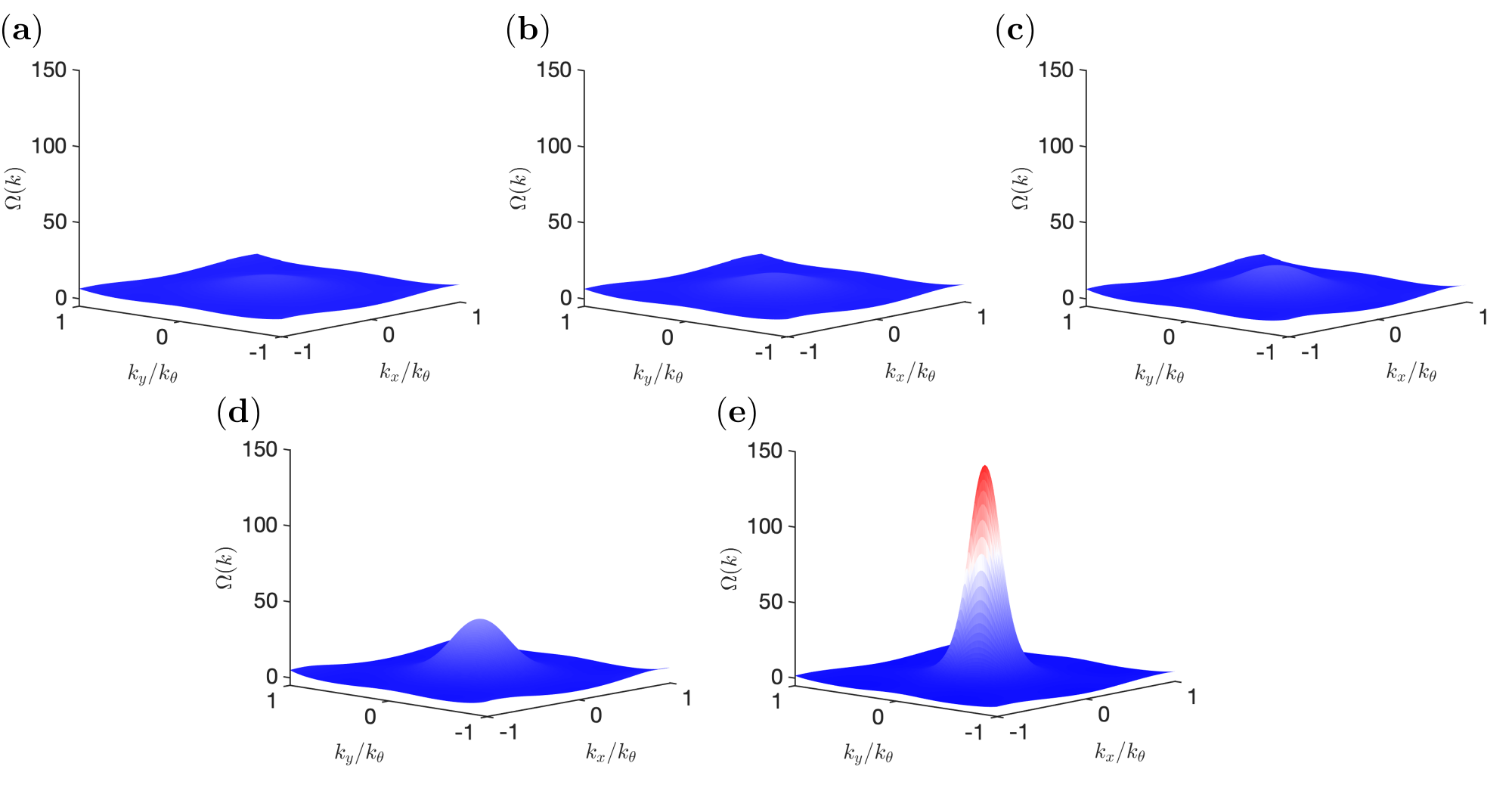}
    \caption{Berry curvature  \cite{Ledwith} for the $\sigma_z = +1$ band of magic angle graphene for different values of $\kappa$.  Panels $\bf{(a)-(e)}$ have $\kappa = 0, \, 0.2 , \, 0.4,\, 0.6,\, 0.8$ respectively.}
    \label{fig:berryC}
\end{figure}

Our situation of degenerate Chern bands seems extremely close to that of quantum Hall ferromagnetism with two important differences.  The first is that there is an extra time reversed copy with opposite Chern numbers.  The second is that these flat bands are not precisely the same as the LLL.  How can we understand and quantify the latter distinction? The coarse indicators of flat dispersion and Chern number are the same.  

Before we answer this question, we make a side comment on how $\kappa > 0$ affects the bands. Because chiral symmetry is no longer present, there is no longer a basis where the wavefunctions are completely sublattice polarized.  Additionally, there is a small dispersion that one may choose to diagonalize. However, one may also choose the basis where the sublattice operator $\Gamma_{ab}(\bk) = \bra{\tilde{u}_{a\bk}} \sigma_z \ket{\tilde{u}_{b \bk}}$ is diagonal. The two bands are exchanged by $C_2 \T$ as before.  Furthermore, $C_2 \T$ symmetry implies that in this basis the non-Abelian Berry connection $\bA_{ab}(\bk) = \bra{\tilde{u}_{a \bk}} \bnabla_\bk \ket{\tilde{u}_{b \bk}} $ is diagonal, we may therefore continue to use sublattice $\sigma_z = \pm1$ to label the two bands, and the two bands continue to have opposite Chern numbers in this basis. The band dispersion is off diagonal in this sublattice basis. The Berry curvature of the $\sigma_z = +1$ band is shown in Figure \ref{fig:berryC} for various values of $\kappa$.  At large $\kappa$ there is a sharp increase in the Berry curvature at the $\Gamma$ point due to approaching band crossings with the remote bands.

\subsubsection{Quantum Geometry of the Chiral Flat Bands:} To understand the similarities and differences between the chiral flat TBG bands and Landau levels we review how to characterize the geometry of a band. We begin by considering the Bloch overlaps, or form factors, $\braket{u_{\bk_1}}{u_{\bk_2}}$ that appear in any physical quantity that is indifferent to the ``internal" indices of the wavefunctions: in the case of chiral TBG, this is layer and the position $\br$ within a unit cell.  There are two important aspects of these overlaps: their phase
and magnitude.  The bare phase is not gauge invariant, but if we consider the phase of an overlap that traces a small loop in $\bk$-space we obtain a gauge invariant quantity from which we may extract the Berry curvature:
\begin{equation}
    \begin{aligned}
        \Omega(\bk) & = \lim_{q \to 0} 
    q^{-2}\,{\Im} P_q(\bk) \\
        P_q(\bk) & =
        \braket{u_{\bk - \frac{q\hat{x}}{2} - \frac{q\hat{y}}{2} }}{u_{\bk + \frac{q\hat{x}}{2} - \frac{q\hat{y}}{2}}}
        \braket{u_{\bk + \frac{q\hat{x}}{2} - \frac{q\hat{y}}{2} }}{u_{\bk + \frac{q\hat{x}}{2} + \frac{q\hat{y}}{2}}} \\
        & \quad \times
        \braket{u_{\bk + \frac{q\hat{x}}{2} + \frac{q\hat{y}}{2} }}{u_{\bk - \frac{q\hat{x}}{2} + \frac{q\hat{y}}{2}}} 
        \braket{u_{\bk - \frac{q\hat{x}}{2} + \frac{q\hat{y}}{2} }}{u_{\bk - \frac{q\hat{x}}{2} - \frac{q\hat{y}}{2}}} 
\end{aligned}
\end{equation}

However, Berry curvature is not sufficient to describe band geometry on its own. For example, all Landau levels with constant magnetic field have the same Berry curvature $\Omega = \ell_B^2$ but they have very different band geometry with important implications: charge density waves become the ground state at fractional Landau level filling after the first few Landau levels.  The origin of this difference is that there is information contained in the magnitude of the Bloch overlaps as well that the Berry curvature does not take into account. The variation in magnitude corresponds to the quantum distance between $\bk$-points and associated Fubini-Study metric
\begin{equation}
    D(\bk_1, \bk_2) = 1-\abs{\braket{u_{\bk_1}}{u_{\bk_2}}},  \qquad g_{ab}(\bk) = \frac{\partial}{\partial q_a} \frac{\partial}{\partial q_b} D(\bk, \bk + \bq)\big|_{\bq  = 0}.
\end{equation}
For the $n$'th Landau level, where zero is the lowest, the Fubini study metric is
\begin{equation}
g_{ab}(\bk) = \left(n + \frac{1}{2} \right)\ell_B^2\delta_{ab}. 
\end{equation}
The increase relative to the LLL comes from the Laguerre polynomial in the form factor $\abs{\langle u_\bk | u_{\bk + \bq} \rangle} = L_n(q^2 \ell_B^2/2)\exp(-q^2 \ell_B^2/4)$. The node in the form factor due to the Laguerre polynomial helps stabilize charge density wave states in higher Landau levels  \cite{HigherLL}.

The Berry curvature and Fubini-Study metric are tied together in special systems.  One may see this by constructing a gauge invariant complex tensor sometimes referred to as the ``quantum metric" which is schematically written as $\eta = g-\frac{1}{2}i\epsilon \Omega$, where $g$ is the symmetric F-S metric and $\epsilon$ is the antisymmetric matrix. This can be compactly written in terms of the gauge invariant projector $Q(\bk)$ as:
\begin{equation}
    \eta_{ab}(\bk) = \frac{\bra{\partial_{k_a} u_\bk}Q(\bk) \ket{\partial_{k_b} u_\bk}}{\braket{u_\bk}{u_\bk}}, 
    \qquad Q(\bk) = 1- \frac{\ket{u_\bk}\bra{u_\bk}}{\braket{u_\bk}{u_\bk}}.
    \label{qtmmetric}
\end{equation}
The quantum metric is Hermitian and positive semidefinite because it is proportional to a projection of $Q(\bk)$ onto the subspace spanned by the states $\ket{\partial_{k_x}u_\bk}$ and $\ket{\partial_{k_y}u_\bk}$, and $Q(\bk)$ is manifestly Hermitian and positive semidefinite as a projection operator.  The real and imaginary components of $\eta$ can be seen to correspond to the Fubini Study metric and Berry curvature respectively
\begin{equation}
    \Omega = - \text{Im}\, \eta_{xy} =  \text{Im}\, \eta_{yx}, \qquad g_{ab} = \text{Re}\, \eta_{ab}
\end{equation}
and the fact that $\eta$ is positive semidefinite, $\det \eta \geq 0$, implies the determinant inequality that $\det g( \bk ) \geq \abs{\Omega(\bk)}^2/4$. This inequality is saturated for all two band systems, and for special multiband systems. Equality in the former case ensues since  the Hilbert space of the two band system corresponds to points on the sphere $S^2$,  for which both the Berry curvature $\frac12|\Omega| $ and the `volume form' $\sqrt{det\,g}$ amount to measuring the surface area.}

In fact,  by using the inequality between the arithmetic mean and the geometric mean of the eigenvalues of $g$ we can derive:
\begin{equation}
\frac12 |{\rm tr}\,g  (\bk )|\geq \sqrt{\det g( \bk )} \geq \frac12\abs{\Omega(\bk)}
    \label{Eqn:G_Omega}
\end{equation}

 A stronger condition that very few systems satisfy, called the Trace condition, is obtained if we additionally require that the eigenvectors of $g(\bk)$ are the same for all $\bk$.  This is not implied by any rotation symmetry, as the rotation symmetry changes the value of $\bk$. In this case, we may simultaneously diagonalize $g(\bk)$ at all values of $\bk$ and obtain:
\begin{equation}
    g_{ab}(\bk) = \frac{1}{2}\abs{\Omega(\bk)} \delta_{ab}, \qquad \tr g(\bk) = \abs{\Omega(\bk)}.
    \label{trcond}
\end{equation}
In practice, for a system like chiral MATBG, this diagonalization amounts to using the coordinates $k_x, k_y$ instead of, say, the triangular lattice coordinates $k_1, k_2$. One may imagine a more exotic reparameterization of the BZ where the change of basis is $\bk$-dependent. This would allow the metric to be diagonalized at all $\bk$ for any system. However we do not consider transformations like this because the transformed BZ coordinates no longer have a simple interpretation as a Fourier transform of position.

The metric \eqref{trcond} saturates the inequality $\tr g \geq \abs{\Omega(\bk)}$. The trace condition \eqref{trcond} is much harder to satisfy than the analogous determinant condition which it implies given the inequality \ref{Eqn:G_Omega}. We will soon show that \eqref{trcond} is satisfied exactly by the {\em lowest} Landau level (which further satisfies that $\Omega(\bk)$ independent of $\bk$) as well as chiral TBG. It is strongly violated for all higher Landau levels. Such a distinction is important because the physics of higher Landau levels is very different from the lowest Landau level, even though all Landau levels are flat and have identical Berry curvature and Chern number.  In particular, most fractional quantum Hall states are observed in the lowest Landau level, and none are observed for the
second and higher Landau levels. 

Notably, if the Bloch states $\ket{u_\bk}$ can be chosen to be holomorphic in $k = k_x + i k_y$, then \eqref{trcond} is satisfied.  The reason can be seen directly from \eqref{qtmmetric} upon replacing $\partial_{k_x} u_k = (\partial_k + \ov \partial_k)u_k = \partial_k u_k$ and similarly $\partial_{k_y} u_k = i \partial_k u_k$. We then obtain
\begin{equation}
    \eta(\bk) \propto \begin{pmatrix} 1 & i \\ -i & 1 \end{pmatrix}
\end{equation}
from which \eqref{trcond} follows. This is the reason why the LLL Bloch states satisfy \eqref{trcond}, and in that case continuous magnetic translation symmetry additionally implies that $\Omega(\bk)$ is independent of $\bk$.  

For chiral TBG there is no continuous magnetic translation symmetry, but the wavefunctions $u_\bk$ in \eqref{chtbg_wfs} are in fact holomorphic in $k$. We ensured this by carefully choosing $f_\bk(z)$, but such a choice was always possible because
the zero mode operator $D$ only contains antiholomorphic derivatives.  Indeed, acting on $u_\bk$ the zero mode equation has the form
\begin{equation}
    \begin{pmatrix} -2i\ov \partial - k & \alpha U(\br) \\ \alpha U(\br) & -2i \ov \partial -k - q_1 \end{pmatrix} u_\bk(\br) = 0.
    \label{holo_u}
\end{equation}
Because $u_\bk$ is a zero mode of an operator that is independent of $\bar k$, it may also be chosen to be independent of $\bar k$. The situation is the same for the LLL.

\begin{figure}
   \centering
\includegraphics[width = \textwidth]{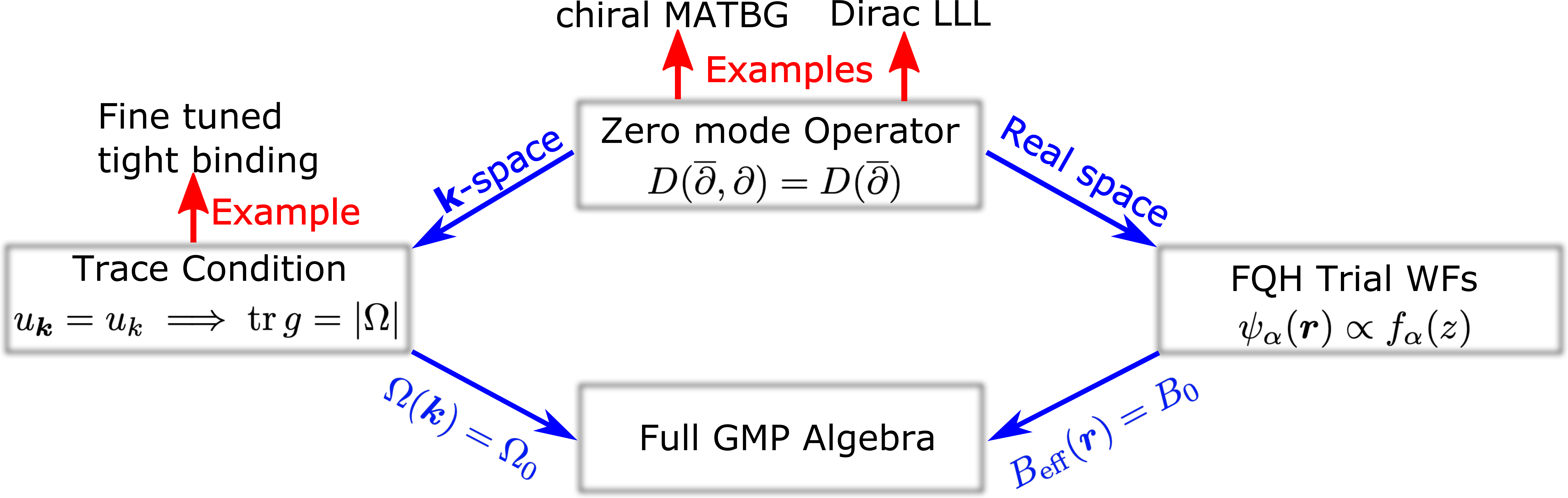}
\caption{Flowchart for results of importance to FCI physics. Both the chiral TBG flat band and Dirac LLL (for a possibly inhomogeneous magnetic field) are zero modes of an operator that only contains antiholomorphic derivatives. The real space dependence of the Bloch states thus has a holomorphic factor which may be used to write fractional quantum hall (FQH) trial wavefunctions.  The periodic wavefunctions $u_\bk$ are only dependent on $k = k_x + i k_y$ (see the discussion around \eqref{holo_u}) which implies the ``trace condition" \eqref{trcond}. The trace condition can also satisfied by fine tuned tight binding models. The trace condition together with flat Berry curvature implies the GMP algebra which is satisfied by the LLL with a constant magnetic field.}
    \label{fig:FCI}
\end{figure}

In fact, the chiral TBG wavefunctions are closely related to LLL wavefunctions. Indeed we may write them in the form
\begin{equation} 
    \psi_{\bk}(\br)  = e^{2\pi i k_1 z/a_1} \vartheta_1\left(\frac{z-z_0}{a_1} - \frac{k}{b_2} \bigg| e^{i \phi}\right) e^{-K(\br)} \begin{pmatrix} v_1(\br) \\ v_2 (\br) \end{pmatrix}
\end{equation}
with
\begin{equation}
    e^{-K(\br)} = \frac{\norm{\psi_{K}(\br)}}{\abs{\vartheta_1\left(\frac{z-z_0}{a_1} | e^{i \phi} \right)} } ,
\end{equation}
as the analogue of the Landau gauge factor $e^{-y^2/2\ell_B^2}$ and $v_i(\br) = \psi_{Ki}(\br)/\norm{\psi_K(\br)}$ are components of a normalized layer-spinor, where here and above $\norm{\psi_K(\br)}$ is the norm of the layer spinor $\psi_K(\br)$. The normalized layer spinor is independent of $\bk$ and so drops out of all Bloch overlaps; it has no influence on band geometry. The wavefunction obtained by stripping off the layer spinor
\begin{equation}
    \tilde{\psi}_\bk(\br) = e^{2\pi i k_1 z/a_1} \vartheta_1\left(\frac{z-z_0}{a_1} - \frac{k}{b_2} \bigg| e^{i \phi}\right) e^{-K(\br)} 
\end{equation}
is the Bloch state for a Dirac particle in an inhomogeneous, Moir\'{e} periodic, magnetic field $B(\br) = \frac{\hbar}{e} \nabla^2 K(\br)$. See \cite{Ledwith} for further details; Ref. \cite{wang2021} additionally shows that such a LLL description holds for all $C=1$ continuum models that satisfy the trace condition.

We may use this understanding to deduce the zero modes of chiral twisted bilayer graphene in an additional \emph{real} magnetic field. 
Let us add an external magnetic field to chiral TBG via 
\begin{equation}
    -i \ov \partial \to -i \ov \partial - e \ov A^{\rm ext}, \qquad \bnabla \times \bA^{\rm ext}(\br) = \bB^{\rm ext}(\br)
\end{equation}
where $\ov A^{\rm ext} = \frac{1}{2}\left(A^{\rm ext}_x - i A^{\rm ext}_y\right)$. Our new zero mode operator becomes $D - 2e \ov A^{\rm ext}$. We define $K^{\rm ext}(\br)$ such that $i \ov \partial K^{\rm ext} = e \ov A^{\rm ext}$. Then we find that we may construct zero modes by multiplying the $B^{\rm ext} = 0$ chiral TBG wavefunctions by $e^{-K^{\rm ext}}$:
\begin{equation}
\begin{aligned}
    \left(D - 2e \ov A^{\rm ext}\right)\left(f(z) e^{-K} e^{-K^{\rm ext}} \begin{pmatrix} v_1(\br) \\ v_2(\br) \end{pmatrix}\right) = e^{-K^{\rm ext}} D \left(f(z)e^{-K} \begin{pmatrix} v_1(\br) \\ v_2(\br) \end{pmatrix}\right) = 0.
\end{aligned}
\end{equation}
These wavefunctions are precisely those of a Dirac particle moving in an external magnetic field $B(\br) + B^{\rm ext}(\br) = \frac{\hbar}{e^2}\nabla^2(K + K^{\rm ext})$, together with the state-independent layer space spinor. See also similar discussions in Refs \cite{Popov2020,sheffer2021}.

% If we choose $B^{\rm ext}(\br)$ to be moir\'{e} perioidic and to have an integer number of flux quanta $n_\Phi$ per moir\'{e} unit cell then we may still use commuting translation operators $T(\ba_{1,2})$ to define Bloch state
% Consider if we thread $n_\Phi$ flux quanta per moir\'{e} cell then the wavefunctions of our system will correspond to those of a Dirac particle moving in a periodic magnetic field with $n_\Phi \pm \sigma_z \tau_z$ flux quanta per unit cell multiplied by the usual $\bk$-independent layer spinor. There will correspondingly be $n_\Phi \pm \sigma_z \tau_z$ perfectly flat bands. 

\subsubsection{Chiral TBG and Fractional Chern Insulators:}
 In this section we argue that  TBG in the chiral limit admits a FCI ground state at fractional filling. We first discuss the experimental setting and background. In certain experimental samples \cite{DavidGG, AndreaYoung} twisted bilayer graphene hosts a quantum anomalous Hall effect at $\nu = 3$. In these samples the graphene layers are thought to be nearly aligned with the hBN substrate. When aligned, the hBN substrate gives the TBG bands a sublattice potential \cite{Zhang19,Bultinck2019,XieMacdonald} that splits the two flat bands in each valley into a $C = +1$ and $C = -1$ band. Then at $\nu = 3$ the system spontaneously breaks time reversal symmetry and chooses one Chern band in one valley to leave unoccupied: the filled bands then have a net Chern number as well and give rise to an anomalous Hall effect. One may hope to obtain a fractional QAHE effect as well, i.e. an FCI, by fractionally filling the unfilled Chern band.

The special quantum geometry and holomorphicity of chiral TBG has important implications for FCI physics. In particular, we may use the real space holomorphicity of the wavefunctions, or the relationship to LLL wavefunctions, to directly write down trial fractional quantum Hall many body wavefunctions. For example, we may write down the Laughlin state at filling $1/m$ 
\begin{equation}
    F(z_1, \cdots, z_n) \propto \prod_{i < j} (z_i - z_j)^m,
\end{equation}
where we used that because the single particle states only depend on a choice of holomorphic function, the many body states are also proportional to a holomorphic function of $n$ complex coordinates.
See Ref.\,\cite{Ledwith} for an explicit construction of the $m$ degenerate Laughlin states for chiral TBG on a real space torus that braid into each other upon inserting fluxes through the cycles (mirroring the discussion in Ref.\,\cite{Haldane1985} for the LLL case).

As in the QHE case the Laughlin state is expected to be a ground state for short range interactions. The argument that is often made for this uses rotation symmetry to project the interaction to the LLL in the symmetric gauge basis for two body states (the Haldane psuedopotentials \cite{Haldane1983}). However, the relative angular momentum here essentially labels the power law describing how fast the wavefunction vanishes as two particles approach each other. The latter makes sense even without rotational symmetry. Thus, we may argue for fractional quantum Hall states even without rotation symmetry (see Ref. \cite{TrugmanKivelson} where this argument was made some time ago for the lowest Landau level). 

We begin by expanding the interaction potential into ``pseudopotentials", written in a real space basis. To understand this expansion, we note that the potential $V(\br)$ will always be integrated against some matrix element or probability density of states in the chiral TBG band; call this function $\Phi(\br)$. The function $\Phi(\br)$ varies on the scale of the moir\'{e} length $a_M$. Let us compute
\begin{equation}
    \int d^2 \,\br V(\br) \Phi(\br).
\end{equation}
Consider the limit where the interaction potential has a range $\ell \ll a_M$. Then to a good approximation we may expand in derivatives of $\Phi$, since only the values of $\Phi$ near $\Phi=0$ matter. We also may assume without loss of generality that $\Phi$ has circular symmetry, $\Phi(\br) = \Phi(r)$, because it is integrated against $V(\br)$ and $V(\br)$ is circularly symmetric. We then use the Taylor-like expansion $\Phi(\br) = \sum_{n=0}^\infty c_n^{-1}(\nabla^{2n} \Phi)(\br = 0) r^{2n}$, where $c_n = (\nabla^{2n} r^{2n})|_{r=0} = 1,4,28,288, \ldots$. This gives
\begin{equation}
    \int d^2 \br\, V(\br) \Phi(\br) = \sum_{n=0}^\infty c_n^{-1} ((a_M\nabla)^{2n} \Phi)(0) \int d^2 \br \left(\frac{r}{a_M}\right)^{2n}V(\br) 
\end{equation}
Here we have inserted factors of $a_M$ to effectively non-dimensionalize $r$ because $\Phi$ varies on the scale of $a_M$. The above result may also be interpreted as coming from the ``pseudopotential" expansion
\begin{equation}
    V(\br) = \sum_{n=0}^{\infty} v_n (a_M \nabla)^{2n} \delta(\br), \qquad v_n = \frac{1}{c_n}\int d^2 \br \left(\frac{r}{a_M}\right)^{2n} V(\br) \sim \left(\frac{\ell}{a_M}\right)^{2n} v_0,
    \label{pseudo_expand}
\end{equation}
through the use of integration by parts.

For $\ell \ll a_M$ the coefficients $v_n$ rapidly decrease with $n$; we therefore imagine restricting the above series to $n \leq m_*$. We now argue that the Laughlin ground state pays zero energy to this truncated interaction for $m>m_*$ and is therefore the ground state because the interaction term is positive semidefinite. The energy expectation value in the Laughlin state $\Psi_m(\br_1,\cdots \br_n)$ is
\begin{equation}
\begin{aligned}
    E & = \sum_{i<j} \frac{1}{2}\int d^2 \br_i d^2 \br_j\, V(\br_i - \br_j)\abs{\Psi_m(\br_1,\cdots \br_n)}^2\\
    & = \sum_{i<j} \sum_{n=0}^{m_*} \int d^2 \br_i d^2 \br_j\,  v_n \left((a_M \nabla)^{2n} \delta(\br_i - \br_j)\right) \abs{\Psi_m(\br_1,\cdots \br_n)}^2.
\end{aligned}
\end{equation}
We now integrate by parts and use that $\abs{\Psi_m}^2 \sim \abs{\br_i - \br_j}^{2m}$ as $\br_i\to \br_j$. Noting that $(\nabla^{2n} r^{2m})|_{r=0} = 0$ for $n<m$, we find that $E = 0$.

As we have discussed, the momentum space band geometry of chiral TBG is also the same as that of the LLL if we allow for inhomogeneous magnetic fields. The condition \eqref{trcond} and the mostly flat Berry curvature have been recognized in the FCI literature as important for reproducing the Girvin-MacDonald-Platzmann (GMP) algebra satisfied by LLL density operators \cite{Parameswaran2012,Roy2014,Parameswaran2013} and magnetic translation symmetry \cite{Claassen2015}. See Fig. \ref{fig:FCI} for a flowchart reviewing the logic of this section as it relates to FCI physics. Finally, we have not delved into the reports of numerical observation of FCI phases in magic angle graphene for which the reader is referred to Refs.~\cite{Repellin,Bergholtz20,Lauchli21}, see also Ref.~\cite{Andrews2020}.

\section{Interaction Effects at Integer Fillings}
\begin{shaded}
In this section we begin by recounting the  simplest electronic  ferromagnet, the quantum Hall system at unit filling in the lowest Landau level. We then describe how twisted bilayer graphene may be thought of as a {\em pair} of generalized quantum Hall ferromagnets, with opposite effective magnetic fields. A large $U(4) \times U(4)$ symmetry rotates the flavors within each quantum Hall system independently. It is broken down to the diagonal $U(4)$ by dispersion which couples the two sectors, and finally the incomplete sublattice polarization lowers the enlarged symmetry to the physical symmetry (i.e. $U(2) \times U(2)$ symmetry of independent spin and charge rotations in each valley). This framework will enable us to understand many integer filling states and will form the basis of the skyrmion mechanism of superconductivity discussed in the next section.
\end{shaded}

\subsection{From Quantum Hall Ferromagnetism to TBG:}
\label{sec:QHFM}
Let us briefly review some basic facts of quantum Hall ferromagnetism (an excellent reference for the following discussion is Girvin's Les-Houches Lecture Notes \cite{Girvin-LesHouchesNotes}). The setup is as follows. Electrons in the lowest Landau level are at filling $\nu=1$. Naturally an integer quantum Hall state is formed, but electrons also have a two fold spin degeneracy that needs to be resolved.  
%The question that remains is about the spin degeneracy. 
Athough Zeeman coupling splits the spin degeneracy, this is often a weak effect, particularly if the $g$ factor is small. In fact, interactions lead to a spontaneous ferromagnet even in the absence of Zeeman field. Furthermore the expected energy gaps are set by the interaction strength, and much larger than the typical Zeeman splitting. The emergence of ferromagnetism for a generic repulsive interaction can be seen as a consequence of Hund's rule where the electrons choose to fill a set of available orbitals in a way that maximizes the total spin \cite{Moon}. The symmetry of the spin part of the many body wavefunction implies the antisymmetry of the orbital part which minimizes the electrostatic repulsion between electrons.

The spontaneous ferromagnet has Goldstone modes (ignoring the Zeeman coupling) characterized by a stiffness. Most interestingly, a consequence of the Chern number of the Landau level is that there is an entanglement between spin and charge degrees of freedom. Topological textures of spin (skyrmions) carry  electric charge \cite{Sondhi93}. This can be understood by imagining an electron moving adiabatically through the spin texture such that its spin is tied to the local spin. This can be shown to generate a Berry phase equivalent to the effect of a fictitious magnetic field given by $b(\br) = \frac{\Phi_0}{4\pi} \bn \cdot (\partial_x \bn \times \partial_y \bn)$ \cite{Girvin-LesHouchesNotes}. Using the Streda relation between charge and flux in a gapped state $\frac{d\rho}{dB} = \sigma_{xy} = C \Phi_0^{-1}$, this immediately gives rise to a charge density $\delta \rho = \frac{C}{4\pi} \bn \cdot (\partial_x \bn \times \partial_y \bn)$ which integrates to a total charge $Q = C W[\bn]$ where $W[\bn]$ is the winding number of the skyrmion texture.

Let us briefly describe this below and transcribe this to our more complex, but similar setting of magic angle TBG. The most important differences will be, first, that in addition to spin, valley quantum number will also be present. Second, given the obvious time reversal symmetry, a mirror sector with opposite magnetic field or equivalently, opposite Chern number will be present. This is shown in the table below. 

\begin{table}[h]
\small
\centering
 \begin{tabular}{c | c|  c} 
 \hline
   & QHFM &  TBG  \\
 \hline\hline
 Ground State  & Ferromagnet at $\nu=1$ & Flavor Order at $\nu={\rm integer} $ \\ 
 Single particle excitations  & Gapped & Gapped    \\ 
Collective modes & Spin waves  & Flavor waves \\ 
$\frac{2E_{\rm skyrmion}}{\Delta_{\rm PH}} = \frac{8\pi \rho}{\Delta_{\rm PH}}$ & $\frac{1}{2}$ & depends on details (Fig.~\ref{fig:SkPHTBG}) \\
 \multirow{2}{*}{Effective Theory} & ${\mathcal L}_C[\bn]={\mathcal L}_{\rm B}+\frac\rho2(\nabla \bn)^2$   & \multirow{2}{*}{${\mathcal L}_+[\bn_+]+{\mathcal L}_-[\bn_-]+J\bn_+\cdot \bn_-$} \\ 
 & $+i \frac{\epsilon^{\mu\nu\lambda}}{8\pi}CA_\mu \bn\cdot \partial_\nu  \bn\times \partial_\lambda \bn $ & \\
 \hline \hline
 \end{tabular}
 \caption{\small Comparison between Quantum Hall ferromagnetism and twisted bilayer graphene (TBG). In the last row,  ${\mathcal L}_{\rm B}$ is the Berry phase for spin 1/2, and the last row and column describes the effective theory for a simplified `spinless' TBG. }
\end{table}

There are some important consequences of the contrast from the quantum Hall case, i.e. arising from the fact that here we have a pair of quantum Hall ferromagnets with opposite Chern number. These are   coupled together by a superexchange interaction $J$, which locks them into an antiferromagnetic configuration. We will see that corresponding to the Skyrmions of the integer quantum Hall effect, in this TBG model the skyrmions will carry twice the charge and correspond to Cooper pairs. Also, an important distinction is that there is an effective time reversal symmetry here that ensures skyrmions do not see a net Lorentz force; the Lorentz force is always odd under time reversal symmetry (it is proportional to a magnetic field). This is in contrast to IQH skyrmions that effectively experience a strong Lorentz force from the large effective magnetic field from the ferromagnetic moments.

\subsection{Projection onto the flat bands}

Let us now turn our attention back to twisted bilayer graphene and discuss the effect of adding the Coulomb interaction to the flat bands. Our main assumption is that the gap to the remote bands $\Delta$ is much larger than the scale for the Coulomb interaction which is given by $V_0 = \frac{e^2}{4\pi \epsilon \epsilon_0 L_M}$ where $\epsilon$ is the relative permittivity and $L_M$ is the Moir\'e length. For parameters typical to TBG, $V_0$ is around 10-20 meV whereas the gap $\Delta$ is close to 100 meV in the chiral limit (in the realistic model, the gap is reduced to about 40 meV). This is similar to projecting into the lowest Landau level.   

To consider the interacting problem, we need to take into account the existence of two graphene valleys $K$ and $K'$ which in addition to spin degeneracy yields a fourfold flavor degeneracy. Combined with the two-fold degeneracy of the flat bands, this means we have 8 flat bands in total. These flat bands are labelled by a spin $s = \uparrow/\downarrow$, valley $\tau = K/K'$ and sublattice $\sigma = A/B$ indices. We will find it convenient to combine these 3 indices into an 8 component index $\alpha = (s, \tau, \sigma)$ labeling the states within the 8-dimensional space of flat bands at a given momentum. Our task is to understand the nature of the ground state at various filling $\nu$. By convention $\nu=0$ refers to the charge neutrality point and $\nu = \pm 4 $ refers to full and empty bands. We will be particularly interested in integer fillings, where an insulator can be stabilized even without enlarging the Moir\'e unit cell. The following analysis follows closely Ref.~\cite{KIVCpaper}. %\AV{Include the figure from soft modes paper with partially filled nearly flat bands?}

The interacting Hamiltonian projected onto the flat bands have the form
\beq
\H = \sum_{\alpha, \bk \in \text{MBZ}} c_{\alpha,\bk}^\dagger h_{\alpha,\beta}(\bk) c_{\beta,\bk} + \frac{1}{2A} \sum_\bq V_\bq \delta \rho_\bq  \delta \rho_{-\bq}, \qquad \delta \rho_\bq = \rho_\bq - \bar \rho_\bq
\label{Hint}
\eeq
Here $A$ denotes the area, $c_{\alpha,\bk}$ denotes the annihilation operator for the electron belonging to the flat band $\alpha$ at momentum $\bk$ in the Moir\'e Brillouin zone. $\delta \rho_\bq$ denotes the Fourier components of the density operator projected onto the flat bands with the background density at charge neutrality subtracted. It can be written explicitly by defining the form factor matrix
\beq
[\Lambda_\bq(\bk)]_{\alpha,\beta} = \langle u_{\alpha, \bk}| u_{\beta, \bk + \bq} \rangle
\eeq
Here, $|u_{\alpha,\bk} \rangle$ denotes the periodic Bloch wavefunctions (which differ from the Bloch wavefunctions by the factor $e^{i \bk \cdot \br}$) for the flat bands. Using $\Lambda$, we can write the density operator $ \rho_\bq$ as
\beq
\rho_\bq = \sum_{\alpha,\beta,\bk} c_{\alpha, \bk}^\dagger [\Lambda_\bq(\bk)]_{\alpha,\beta} c_{\beta,\bk + \bq} = \sum_{\bk} c_{ \bk}^\dagger \Lambda_\bq(\bk) c_{\bk + \bq}, \qquad \bar \rho_\bq = \frac{1}{2} \sum_{\bk, \bG} \delta_{\bq, \bG} \tr \Lambda_\bG(\bk)
\label{rhoq}
\eeq
Note here that $\bk$ lies in the Moir\'e Brilluoin zone, $\bq$ is unrestricted, and $\bG$ is a reciprocal lattice vector. In the second equality for $\rho_\bq$ above, we used a more compact matrix notation where $c_\bk$ is an 8-component column vector and $\Lambda_\bq(\bk)$ is an $8 \times 8$ matrix. 

\begin{figure}
    \centering
    \includegraphics[width = 0.5\textwidth]{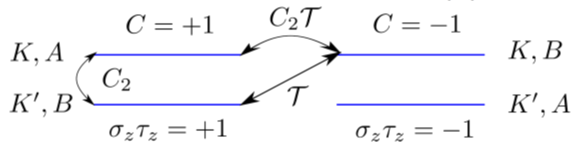}
    \caption{Schematic illustration of the Chern sectors for the flat bands in the spinless limit labelled by valley $K/K'$ and sublattice $A/B$ with the action of time-reversal $\T$, twofold rotation $C_2$ and the combination of the two $C_2 \T$ illustrated. Figure adapted from Ref.~\cite{KIVCpaper}.}
    \label{fig:SpinlessSym}
\end{figure}

Using symmetries in the chiral limit, (see Appendix \ref{Sec:twoparticleSymmetries} for the detailed analysis), we obtain a very simple form of the form factor matrix:
\beq
\Lambda_\bq(\bk) = F_\bq(\bk) e^{i \Phi_\bq(\bk) \sigma_z \tau_z}
\eeq
where $F_\bq(\bk)$ and $\Phi_\bq(\bk)$ are scalars. Note that the form factor of a given flat band only depends on the combination $\sigma_z \tau_z$ which corresponds to the Chern number. This can be understood from the fact that in addition to being diagonal in  spin and valley, bands in the same valley with opposite Chern number live on opposite sublattices.  This leads to the form above, on further noticing that Chern number flips under exchanging sublattice (under $C_2 \T$ symmetry), or valley (under $\T$) but remains invariant under flipping both (under $C_2$ symmetry). This illustrated in Fig.~\ref{fig:SpinlessSym} where the spin degree of freedom is suppressed since it does not play a role in understanding the action of other symmetries (this arises since the problem has a full $\SU(2)$ spin rotation invariance).

The non-interacting part $h_{\alpha,\beta}(\bk)$ is naively obtained by projecting the chiral BM Hamiltonian onto the flat bands
\beq
h_{\alpha,\beta}(\bk) = \langle u_{\alpha,\bk}| H_{\rm BM}(\bk) | u_{\beta, \bk} \rangle
\label{hHBM}
\eeq
which suggests that it vanishes identically at the magic angle. In reality, the process of projecting onto the flat band also generates corrections coming from the interaction with the remote bands that contribute to the single particle dispersion $h_{\alpha,\beta}(\bk)$ \cite{XieMacdonald, Repellin, Shang}. While a detailed calculation of these corrections is beyond the scope of these notes, we can deduce the general form of $h_{\alpha,\beta}(\bk)$ and of the form factors $\Lambda_\bq(\bk)$ based on symmetry alone. See Appendix \ref{Sec:AppendixSymmetries} for a complete discussion of the symmetries of chiral TBG.

% It is instructive to see how these form factor look if we replace the TBG flat bands with lowest Landau level wavefunctions. In this case, the parameters $F_\bq(\bk)$ and $\Phi_\bq(\bk)$ as simply given by
% \beq
% F_\bq(\bk) = e^{-\frac{l_B^2}{2} \bq^2}, \qquad \Phi_\bq(\bk) = \frac{l_B^2}{2} \bk \wedge \bq
% \eeq
% where $l_B$ is the magnetic length. \AV{I thought $F = e^{-q^2l_B^2/4}$ also how did you obtain $\Phi$?}
% \EK{I will add a bit more discussion about the structure of the form factors e.g. for small $\bq$, $\Phi_\bq \approx \bq \cdot A_\bk$}

\subsubsection{Enlarged Symmetry}
Let us now restore the spin degree of freedom and consider the symmetry of the full problem. The simple structure of the form factor implies that the interaction term in the Hamiltonian (\ref{Hint}) has a large symmetry.  To see this, we note that the density operator is invariant under any unitary rotation
\beq
c_\bk \mapsto U c_\bk, \qquad [U, \sigma_z \tau_z] = 0
\eeq
where as before $c_\bk$ is an 8-component column vector which means that $U$ is an $8 \times 8$ matrix. The condition that $U$ commutes with $\sigma_z \tau_z$ simply means we restrict ourselves to unitary rotations which conserve the Chern number i.e. which take place within the same Chern sector. Since there are four bands within each Chern sector $\pm 1$, this leads to a $\U(4) \times \U(4)$ symmetry for the interaction term.

To see how this symmetry is reduced in the presence of the single particle term $h(\bk)$, we need to see how the different symmetries restrict its form. Note that in practice obtaining $h(\bk)$ through direct band projection underestimates the dispersion; the dispersion is substantially enhanced by interactions. Our main assumption is that, although its value may be renormalized by interactions, the effective dispersion has the same symmetries as the band projected dispersion. We also choose the dispersion to be compatible with chiral symmetry. First, we can again use spin and valley symmetries to deduce $h(\bk)$ is proportional to $s_0$ and diagonal in valley index. For a single valley, $h(\bk)$ is further restricted by symmetry as described in the Appendix \ref{Sec:oneparticleSymmetries}.  

This leads to the form:
\beq
h(\bk) = h_x(\bk) \sigma_x + h_y(\bk) \sigma_y \tau_z
\label{hk}
\eeq

\begin{figure}
    \centering
    \includegraphics[width = 0.6 \textwidth]{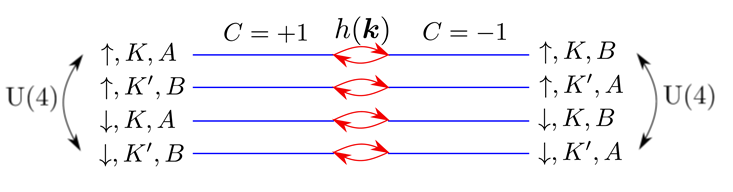}
    \caption{Illustration of the enlarged $\U(4) \times \U(4)$ symmetry corresponding to independent unitary rotation in each Chern sector which is reduced to $\U(4)$ in the presence of tunneling which couples the two sectors together \cite{KIVCpaper}.}
    \label{fig:SpinfulTunneling}
\end{figure}

To understand how this term reduces the $\U(4) \times \U(4)$ symmetry of the interaction, it is useful to introduce a new Pauli triplet $(\gamma_x, \gamma_y, \gamma_z) = (\sigma_x, \sigma_y \tau_z, \sigma_z \tau_z)$. The condition $[U, \sigma_z \tau_z] = 0$ then becomes $[U, \gamma_z ] = 0$ which yields
\beq
U = \left(\begin{array}{cc}
   U_+  & 0 \\
    0 & U_- 
\end{array}\right)
\label{Upm}
\eeq
in the basis where $\gamma_z = \diag(1,1,-1,-1)$. In the $\gamma$ basis, the single particle dispersion takes the simple form $h(\bk) = h_x(\bk) \gamma_x + h_y(\bk) \gamma_y$. Thus, if we want $U$ to also commute with the single-particle part $h(\bk)$, we need to require $[U, \gamma_x] = 0$ in addition which gives $U_+ = U_-$, i.e. $U \propto \gamma_0$. This reduces the $\U(4) \times \U(4)$ of the interaction to a single $\U(4)$. Intuitively, this can be understood by noting that $h(\bk)$ has the form of a tunneling which connects opposite sublattice bands in the same valley (and spin) in the the opposite Chern sector. This requires us to perform the same $\U(4)$ symmetry on both sides to retain the structure of this coupling between the Chern bands thereby reducing $\U(4) \times \U(4)$ to a single $\U(4)$. This is schematically illustrated in Fig.~\ref{fig:SpinfulTunneling}.

Let us now briefly comment on the symmetry reduction once we move away from the chiral limit. As shown in Appendix \ref{sec:DeviationFromChiralLimit}, deviation from the chiral limit means that the wavefunctions are not completely sublattice-polarized. This leads to the appearance of a sublattice off-diagonal contribution to the form factor proportional to $\sigma_x \tau_z$ and $\sigma_y$. To understand the symmetry reduction due to this extra term, it is convenient to introduce the Pauli triplet $(\eta_x, \eta_y, \eta_z) = (\sigma_x \tau_x, \sigma_x \tau_y, \tau_z)$ which commute with $\gamma_{x,y,z}$. We then find that away from the chiral limit, a unitary rotation $U$ leaves the interaction invariant if and only if $[U,\gamma_z] = [U, \gamma_x \eta_z] = 0$. This means that $U_+$ and $U_-$ defined in Eq.~\ref{Upm} are related via $U_+ = \eta_z U_- \eta_z$. This is different from the $\U(4)$ symmetry preserved by $h(\bk)$ which is given by $U_+ = U_-$. In fact, the two conditions are only satisfied if $U_+ = \eta_z U_+ \eta_z$ which corresponds to the physical symmetry group $\U(2) \times \U(2)$ corresponding to independent spin and charge conservation in each valley. We note that deviation from the chiral limit also introduces an extra term $\propto \eta_z = \tau_z$ in the dispersion which turns out to be quite small in practice. The reason for its smallness can be understood perturbatively from the fact that dispersion itself is a small correction compared to the interaction and that deviations from the chiral are small as well. This means that, assuming both perturbations are of similar order, this term appears at the next order in perturbation theory (see discussion in Ref.~\cite{KIVCpaper} for further details about the hierarchy of scales in the problem). %Another way to see the different symmetries is not note that the generators of the $\SU(2)$ symmetry preserved by $h(\bk)$ are $\eta_{x,y,z}$ whereas the generators of the $\SU(2)$ symmetry preserved by the interaction away from the chiral limit are $\eta_{x,y} \gamma_z$, $\eta_z$.

\subsection{Correlated insulators: generalized ``quantum Hall ferromagnets"}
We can now understand the emergence of correlated insulators at integer fillings. For simplicity, we are going to focus on the spinless limit where the interaction has a $\U(2) \times \U(2)$ symmetry which is reduced to $\U(2)$ in the presence of the single-particle dispersion. We will also focus on the case of charge neutrality. The relevance of this limit for TBG can be understood as follows. There is good experimental evidence \cite{CascadeShahal, CascadeYazdani} for spin polarization in the vicinity of half-filling \footnote{Note that due to the presence of $\SU(2)$ spin rotation symmetry in each valley separately, a spin polarized state does not have to be a ferromagnet since we can choose the filled spin independently in each valley.}. Furthermore, this spin polarization takes place at significantly higher temperature compared to those associated with the correlated insulating and superconducting phases. This suggests that the energy scale for this spin ordering (the spin stiffness) is larger than the scales associated with the correlated insulating or superconducting behavior, which justifies projecting out the completely empty/completely full spin sector and focusing on the spinless limit.

Thus, we are going to focus on the spinless model at half-filling i.e. two out of four bands are filled. Let us start by focusing on the interaction term and ignoring the single-particle dispersion. We note that the interaction is a sum of positive semidefinite operators for each momentum $\bq$, thus its spectrum is non-negative. Furthermore, any state which satisfies $\delta \rho_\bq |\Psi \rangle = 0$ for all $\bq$ is a ground state for the interacting part. Let use now write the density operator more explicitly using Eqs.~\eqref{rhoq} and \eqref{Lambdaqk}
\begin{multline}
    \rho_\bq = \sum_\bk F_\bq(\bk) [ e^{i \Phi_\bq(\bk)} (c_{A,K,\bk}^\dagger c_{A,K,\bk + \bq} + c_{B,K',\bk}^\dagger c_{B,K',\bk + \bq}) \\ + e^{-i \Phi_\bq(\bk)} (c_{A,K,\bk}^\dagger c_{A,K,\bk + \bq} + c_{B,K',\bk}^\dagger c_{B,K',\bk + \bq})]
\end{multline}

We see that $\rho_\bq$ consists of four terms which correspond to a momentum space hopping between $\bk$ and $\bk + \bq$ in each of the four bands seperately. Thus, the four terms are gauranteed to vanish if they act on a state where each of the four bands is completely empty or completely full yielding an exact ground state of the interacting Hamiltonian.\footnote{This is only valid if $\bq$ is not equal to a reciprocal lattice vector $\bG$. In this case, the action of $\rho_\bG$ yields a term $c_\bk^\dagger c_{\bk + \bG} = c_\bk^\dagger c_\bk$ that is equal to the filling of electrons at momentum $\bk$. This term is a constant which cancels against the background term $\bar \rho_\bq$ at charge neutrality in $\delta \rho_\bq$ such that $\delta \rho_\bq |\Psi \rangle = 0$ for all $\bq$.}

This yields two possible kinds of ground states at half-filling (2 out of 4). First, we can fill two bands in the same Chern sector yields an anomalous quantum Hall state
\beq
|\Psi_{\rm QAH, +2} \rangle = \prod_{\bk \in \text{mBZ}} c_{A, K, \bk}^\dagger c_{B, K', \bk}^\dagger |0 \rangle, \qquad |\Psi_{\rm QAH, -2} \rangle = \prod_{\bk \in \text{mBZ}} c_{B, K, \bk}^\dagger c_{A, K', \bk}^\dagger |0 \rangle
\eeq
$|\Psi_{\rm QAH, \pm 2} \rangle$ has Chern number $\pm 2$ and map to each other under time-reversal symmetry $\T$ i.e. each of them spontaneously breaks $\T$. Both states are symmetric under the $\U(2) \times \U(2)$ symmetry since unitary rotations among two fully filled or fully empty bands do nothing. The second class of states is obtained by filling one band in each of the $\pm$ Chern sectors leading to a vanishing total Chern number. This can be chosen to be any arbitrary linear combination of the two bands in each Chern sector yielding the state
\beq
|\Psi_{\theta_+,\phi_+, \theta_-, \phi_-} \rangle = \prod_{\bk \in \text{mBZ}} (\cos \theta_+ c_{A, K, \bk}^\dagger + \sin \theta_+ e^{i \phi_+} c_{B, K', \bk}^\dagger) (\cos \theta_- c_{B, K, \bk}^\dagger + \sin \theta_- e^{i \phi_-} c_{A, K', \bk}) |0 \rangle
\eeq
The angles $\theta_\pm$ and $\phi_\pm$ parametrize two unit vectors $\bn_\pm = (\sin \theta_\pm \cos \phi_\pm, \sin \theta_\pm \sin \phi_\pm, \cos \theta_\pm)$. Another way to see this is to start from a simple state, let's say $\theta_\pm = 0$, corresponding to a valley polarized state filling both sublattice bands in the $K$ valley. Since a single band out of 2 is selected to be filled in each Chern sector, this state breaks the $\U(2) \times \U(2)$ symmetry of the interaction. For each Chern sector, the $\U(2)$ symmetry is broken down to $\U(1) \times \U(1)$ corresponding to independent phase rotation in the full or empty band. Thus, acting with $\U(2) \times \U(2)$ on the valley polarized state generates a manifold of states characterized by a unit vector in $S^2 = \frac{\U(2)}{\U(1) \times \U(1)}$ in each of the two Chern sectors.

\begin{figure}
    \centering
    \includegraphics[width = 0.7 \textwidth]{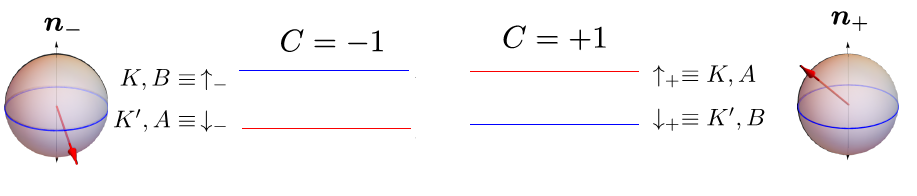}
    \caption{Illustration of the Pseudospin vector with $K, A$ and $K, B$ denoting the up pseudospin direction in the $\pm$ Chern sectors, respectively \cite{SkPaper}. }
    \label{fig:Pseudospins}
\end{figure}

We will call the two unit vectors $\bn_+$ and $\bn_-$ parametrizing the manifold of Chern zero states pseudospins (see Fig.~\ref{fig:Pseudospins}). These can be extracted from the ground state wavefunctionusing the pseudospin Pauli triplet $(\eta_x, \eta_y, \eta_z) = (\sigma_x \tau_x, \sigma_x \tau_y, \tau_z)$ introduced earlier via
\beq
\bn_\pm = \langle \Psi| {\boldsymbol \eta} \frac{1 \pm \gamma_z}{2}| \Psi \rangle
\eeq

It is instructive to look at some special points on this manifold which are invariant under some physical symmetries. First, we notice that the $\U(1)$ valley rotation associated with the conservation of the valley charge acts as a pseudospin rotation in the $x-y$ plane $e^{i \phi \tau_z} = e^{i \phi \eta_z}$. %Time-reversal acts by exchanging the Chern sectors and flipping the $y$-components of the pseudospin $\T: (n_{\pm,x}, n_{\pm, y}, n_{\pm, z}) \mapsto (n_{\mp,x}, -n_{\mp, y}, n_{\mp, z})$ whereas $C_2$ flips the $y$ and $z$ pseudospin components within the same Chern sector $C_2: (n_{\pm,x}, n_{\pm, y}, n_{\pm, z}) \mapsto (n_{\mp,x}, -n_{\mp, y}, n_{\mp, z})$
Thus, if both the pseudospins $\bn_\pm$ point in the $z$-direction, the resulting state would preserve $\U(1)$ valley symmetry. There are two options, either $\bn_+ = \bn_- = \pm \hat z$ corresponding to aligned Ising order between the Chern sectors or $\bn_+ = -\bn_- = \pm \hat z$ corresponding to anti-aligned Ising order between the Chern sector. The former corresponds $\langle \tau_z \rangle \neq 0$ and is obtained by filling the same valley in the two Chern sectors (both sublattice bands in the same valley) leading to a valley polarized state which is invariant under $C_2 \T$ (remember $C_2 \T$ flips sublattice but not valley). The latter corresponds to $\langle \sigma_z \rangle \neq 0$ and is obtained by filling opposite valleys for the two different Chern sectors (or equivalently filling both valleys in only one of the sublattices)  leading to a valley Hall state which is invariant under time-reversal symmetry (remember $\T$ flips valley but not sublattice). In-plane pseudospin order spontaneously breaks $\U(1)$ valley symmetry, but we can still distinguish two cases of XY pseudospin order which is aligned or anti-aligned between the Chern sector. The aligned case corresponds to a non-vanishing expectation value for the order parameter $\langle \sigma_x \tau_{x,y} \rangle = \langle \eta_{x,y} \rangle \neq 0$ which is invariant under time-reversal symmetry whereas the anti-aligned case corresponds to the order parameter $\langle \sigma_y \tau_{x,y} \rangle = \langle \gamma_z \eta_{x,y} \rangle \neq 0$ which is odd under time-reversal, but invariant under a combination of time-reversal and $\pi$ valley rotation $e^{i\pi \tau_z} = i\tau_z$. Since $\T$ anticommutes with $\tau_z$ (it exchanges valleys), the combination $\T' = i \tau_z \T$  acts as a new time-reversal symmetry which squares to $-1$ rather than $+1$, i.e. a Kramers type time-reversal symmetry. A summary of these states is given below in Table \ref{tab:GroundStates}.

\begin{table}[h]
\small
    \centering
    \begin{tabular}{c|c|c}
    \hline \hline
    State & Pseudospin Description & Unbroken Symmetries \\
    \hline
        Valley Polarized $\tau_z$ & Ising Aligned $\eta_z$ 
         & $\U_V(1), C_2\T$ \\
         Valley Hall $\sigma_z$ & Ising Anti-aligned $\gamma_z \eta_z$ & $\U_V(1), \T$ \\
         $\T$-symmetric Intervalley  $\sigma_x \tau_{x,y}$ & XY Aligned $\eta_{x,y}$ & $\T$, \EK{$C_2 e^{i \phi \tau_z}$} \\
         Kramers Intervalley  $\sigma_y \tau_{x,y}$ & XY Anti-aligned $\gamma_z \eta_{x,y}$ & $\T' = i\tau_z \T$, \EK{$C_2 e^{i \phi \tau_z}$} \\
         \hline \hline
    \end{tabular}
    \caption{\small Table summarizing some of the states from the manifold of the Chern zero low-energy states, their corresponding pseudospin description, and their symmetries. The pseudospin order is ferromagnetic within each Chern sector but it can be aligned $\bn_+ = \bn_-$ or anti-aligned $\bn_+ = -\bn_-$ between Chern sectors. Although the intervalley coherent order generically breaks $C_2$ symmetry, it always preserves a combination of $C_2$ and valley rotation $e^{i \phi \tau_z}$ where $\phi$ is related to the angle of the IVC XY order.}
    \label{tab:GroundStates}
\end{table}

\subsubsection{Effect of dispersion}
\label{sec:dispersion}
The analysis of the ground state so far focused on the ground states of the interacting term ignoring the single-particle dispersion $h(\bk)$. As discussed earlier, this term is likely to be non-zero even for the chiral limit at the magic angle and has the general form given by (\ref{hk}) which breaks the $\U(2) \times \U(2)$ symmetry down to a single $\U(2)$. Our main assumption here is that the magnitude of this term is small compared to the interaction scale such that its effect can be included perturbatively leading to a splitting among the ground states of the interaction part.

Before computing this splitting, we will present a simple intuitive argument to understand the effect of this term. $h$ acts as tunneling between opposite Chern (sublattice) bands belonging to the same valley. If we are in a state where both these bands are fully filled or fully empty e.g. in a valley polarized state, then this tunneling term is Pauli blocked and has no effect on the energy of the state. On the other hand, if one band is fully filled and the other is fully empty, this tunneling term is expected to reduce the energy through virtual tunneling. This reduction mechanism is analogous to  superexchange and leads to an energy reduction of the order of $h^2/U$ where $U$ is the interaction scale.

For a state in the manifold of zero Chern insulators described by the pseudospin vectors $\bn_+$ and $\bn_-$, the energy correction due to tunneling can only be a function of $\bn_+ \cdot \bn_-$ due to $\SU(2)$ pseudospin symmetry. Furthermore, if we compute this contribution in second order perturbation theory it can only depend linearly on $\bn_+ \cdot \bn_-$ leading to the energy contribution $\Delta E = J ( \bn_+ \cdot \bn_- - 1)$ where we have added an unimportant constant such that $\Delta E$ vanishes for $\bn_+ = \bn_-$ as expected. The value of $J$ can be computed by taking any of the states with $\bn_+ = -\bn_-$ for which $\Delta E = -2J$. For definiteness, let us take the valley Hall state with $\bn_+ = -\bn_- = \hat z$. Relegating technical details to appendix \ref{app:Tunneling}, we can extract the value of $J$ by computing the second order correction to the energy due to $h$ leading to
\beq
J = \frac{1}{N} \sum_{\bk, \bk'} [h_x(\bk) + i h_y(\bk)] [\H_{\rm eh}]_{\bk,\bk'}^{-1} [h_x(\bk') - i h_y(\bk')],
\eeq
where $\H_{\rm eh}$ is the Hamiltonian for an inter-Chern particle-hole excitation created by the action of $h(\bk)$. Its explicit form is given by \cite{KIVCpaper}
\beq
[\H_{\rm eh}]_{\bk,\bk'} = \frac{1}{A} \sum_\bq V_\bq F_\bq(\bk)^2 [\delta_{\bk,\bk'} - \delta_{\bk',[\bk+\bq]} e^{2 i \Phi_\bq(\bk)}]
\eeq
where $[\bq]$ denotes the part of $\bq$ within the first Moir\'e Brillouin zone. As discussed in Ref.~\cite{KIVCpaper}, this Hamiltonian always has a finite spectral gap due to the topological properties of the bands which guarantees that the perturbation theory is well-defined.

\subsubsection{Deviation from the chiral limit}
\label{sec:devChiral}
Let us now see how the deviation from the chiral limit influences the energies of the states in the ground state manifold. We start by noting that, away from the chiral limit, the wavefunctions are not completely sublattice-polarized leading to sublattice off-diagonal contribution to the form factor matrix proportional to $\sigma_x \tau_z$ and $\sigma_y$ (see appendix \ref{sec:DeviationFromChiralLimit} for details). The extra sublattice off-diagonal contribution to the form factor means that the density operator has a corresponding extra term given by
\begin{align}
\tilde \rho_\bq &= \sum_\bk \tilde F_\bq(\bk) c_\bk^\dagger \sigma_x \tau_z e^{i \tilde \Phi_\bq(\bk) \sigma_z \tau_z} c_{\bk + \bq} \nonumber \\
&= \sum_\bk \tilde F_\bq(\bk) [e^{-i \tilde \Phi_\bq(\bk)} (c_{A,K,\bk}^\dagger  c_{B,K,\bk + \bq} - c_{B,K',\bk}^\dagger  c_{A,K',\bk + \bq}) \nonumber \\
 & \qquad + e^{i \tilde \Phi_\bq(\bk)} (c_{B,K,\bk}^\dagger  c_{A,K,\bk + \bq} - c_{A,K',\bk}^\dagger  c_{B,K',\bk + \bq})]    
\end{align}
This represents hopping between opposite sublattices within the same valley. Including this term, we find that the condition that a state is a ground state for the full interaction term is $\delta \rho_\bq |\Psi \rangle = \tilde \rho_\bq |\Psi \rangle = 0$ for all $\bq$. Note that this is an exact non-perturbative statement that does not assume deviations from the chiral limit are small. This condition is manifestly satisfied by the valley polarized state, but it is also satisfied by the Kramers intervalley coherent state which is generated from the valley polarized state by unitary rotations generated by $\gamma_z \eta_{x,y}$. In fact, we can deduce the form of the energy contribution due to deviation from the chiral limit based on symmetry considerations alone since invariance under the $\SU(2)$ group generated by $\eta_x \gamma_z$, $\eta_y \gamma_z$, and $\eta_z$ restricts the energy to the form 
\beq
\Delta E = \lambda (n_{+,x} n_{-,x} + n_{+,y} n_{-,y} - n_{+,z} n_{-,z} + 1)
\eeq
where we added a constant such that $\Delta E$ vanishes for the valley polarized state as expected. The value of $\lambda$ can be deduced by taking a state which maximizes $\Delta E$, e.g. valley Hall state with $\Delta E = 2 \lambda$. A direct calculation yields
\beq
\lambda = \frac{1}{4A} \sum_\bq V_\bq \langle \Psi_{\rm VH}| \tilde \rho_\bq \tilde \rho_{-\bq} | \Psi_{\rm VH} \rangle = \frac{1}{2A} \sum_\bq V_\bq \tilde F_\bq(\bk)^2
\eeq

\subsubsection{Energy splitting}
Combining the effects of the dispersion and deviation from the chiral limit yields the following energy function for the splitting
\beq
\Delta E[\bn_+, \bn_-] = J \bn_+ \cdot \bn_- + \lambda (\bn_{+,xy} \cdot \bn_{-,xy} - n_{+,z} n_{-,z})
\eeq
This function selects the anti-aligned XY order (the Kramers intervalley coherent state) as the unique ground state.

\subsubsection{Comparison to numerics and experiments}
The main conclusion of our argument is the existence of a large number low-lying insulating states at integer filling $\nu$ which can be thought of as generalized quantum Hall ferromagnets obtained by filling $4 + \nu$ of the $C = \pm 1$ bands shown in Fig.~\ref{fig:SpinfulTunneling}. In general, the manifold of insulating states at a particular filling and Chern number is not parameterized by two pseudospins but instead by two Grassmanians; all the results we obtained here can be directly generalized (see e.g. the appendix of Ref. \cite{SkPaper}). The energy splitting between the various quantum hall ferromagnets is relatively small, $\lesssim$ 1 meV per particle, and can depend sensitively on the sample details e.g. strain, substrate alignment, lattice relaxation, etc. However, the broad conclusion regarding the existence of a large number of Slater determinant generalized ferromagnets has been robustly reproduced in several studies starting from numerical Hartree-Fock studies \cite{XieMacdonald, Shang, KIVCpaper, GuineaHF} which identified several closely competing low-lying states whose competition precisely matches our analysis. Further verification came from less biased numerical methods such as DMRG \cite{Tomo, Parker, VafekDMRG} and exact diagonalization \cite{TBGVI, Potasz} which found that the ground state is almost always very close a Slater determinant state. Later analytical works also reproduced these results \cite{TBGIII, TBGIV}.

\begin{figure}
    \centering
    \includegraphics[width = 0.7 \textwidth]{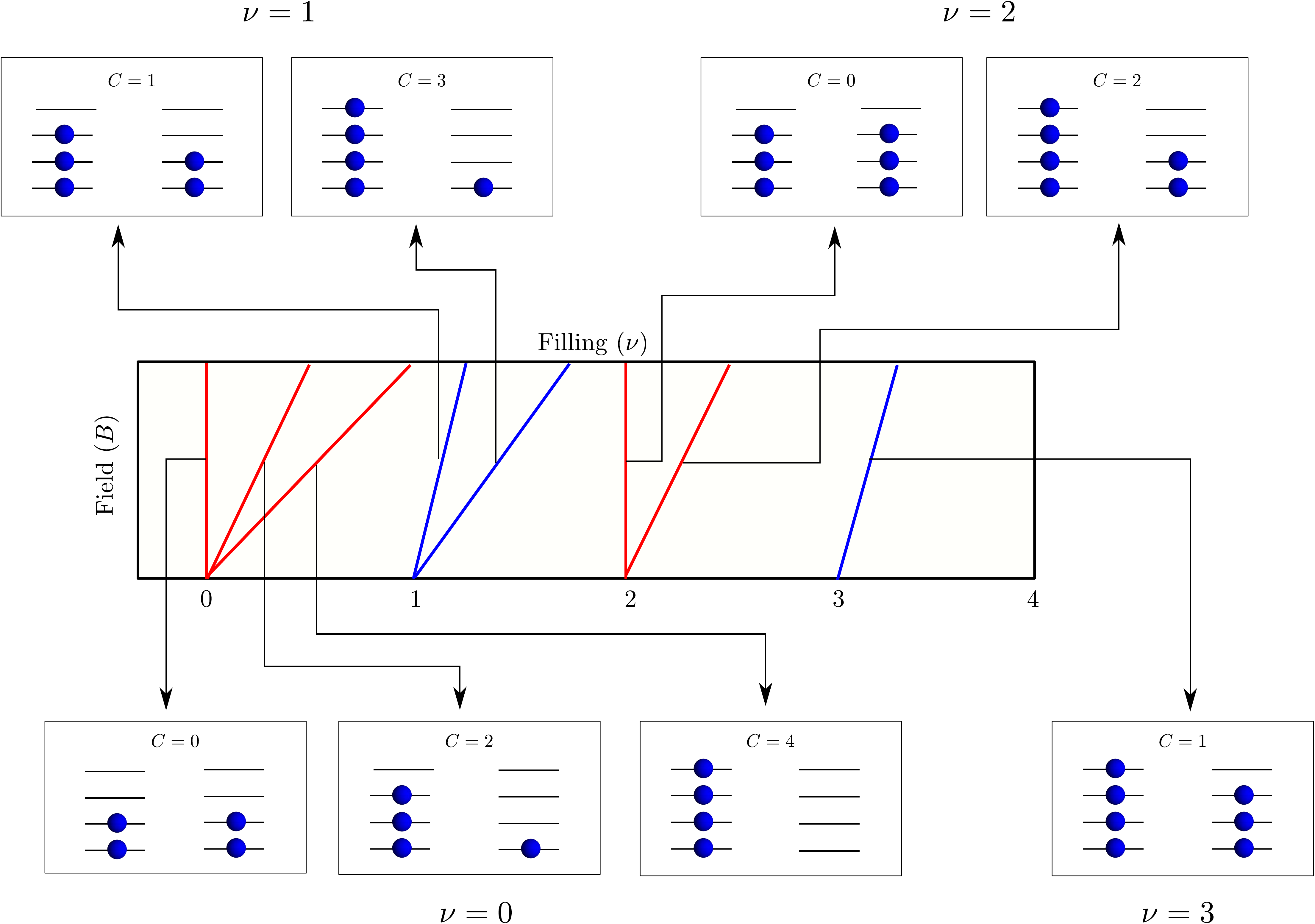}
    \caption{A schematic illustration of the different Chern insulators possible at different integer fillings by filling different bands in the $C = \pm 1$ sectors. Here, we use a Landau fan diagram where the slope of the different lines represent the Chern number and the intercept represents the filling. Figure adapted from Ref.~\cite{SoftModes}. }
    \label{fig:ChernStates}
\end{figure}

Experimentally, the main implication of the existence of a large number of closely competing states is the possibility of inducing phase transitions between them by applying relatively small perturbations. One of the most experimentally accessible such perturbations is the out of plane magnetic field which, due to orbital coupling to the electrons, leads to energy splitting between states with different total Chern number as a function of filling \cite{TBGIV}. The different possible Chern insulators obtained by filling a subset of the $C = \pm 1$ bands at a given filling are shown in Fig.~\ref{fig:ChernStates}. This picture predicts a phase transition to states with finite Chern number with the same parity as $\nu$ whose maximum value is $|C| = 4 - |\nu|$ \cite{KIVCpaper, TBGIV}. This was experimentally verified in several parallel works \cite{ChernCaltech, ChernRutgers, ChernYazdani, ChernYoung} which observed these different Chern insulators upon applying an out-of-plane magnetic field to TBG samples.

\section{Skyrmions and superconductivity}
%\PL{WZW terms: \cite{Christos2020}}
\begin{shaded}
In this section, we discuss a possible scenario for the emergence of superconductivity upon doping the half-filling correlated insulators identified in the previous section. {We first discuss an effective field theory for the manifold of correlated insulators; slowly varying textures of the order parameter field give rise to skyrmions. We then discuss the energetics of skyrmions as well as their kinematics which enables us to obtain a BEC condensation temperature. Finally, we show that skyrmion condensation may be understood through a phase transition of the $\CP^1$ model, and through a large $N$ expansion obtain a complementary estimate of the condensation temperature.}
\end{shaded}

\subsection{Effective field theory}
Our starting point is an effective field theory for the correlated insulator. Such a theory is known as a non-linear sigma model; it describes the subspace of low energy states related to each other by the global $\U(2) \times \U(2)$ symmetry. We begin by restricting ourselves to the manifold of Chern 0 states and exclude the quantum anomalous Hall states with Chern number $\pm 2$. This is motivated by two considerations. First, since the quantum anomalous Hall states do not break any contineous symmetry (they are $\U(2) \times \U(2)$  invariant), there are no low energy degrees of freedom making their effective low energy theory trivial. Second, these states strongly break time-reversal symmetry which is likely to suppress superconductivity. 

To write the field theory for the manifold of Chern zero states, we notice that it is parametrized by a pair of unit vectors $\bn_\pm$ describing the pseudospin magnetization direction in the $\pm$ Chern sector. Let us start by focusing on a single Chern sector. Topologically, this system is equivalent to a quantum Hall ferromagnet with the main difference being the detailed structure of the wavefunctions and the replacement of magnetic translations with regular translations, both features which does not alter the form of the effective field theory (see Refs.~\cite{SkPaper, SkDMRG, SoftModes} for a detailed derivation for the effective field theory for the case of Chern band). Thus, the field theory for a single Chern sector with $C = \pm 1$ is the same as that of a quantum Hall ferromagnet \cite{Girvin-LesHouchesNotes, Moon}
\beq
\L_\pm[\bn] = \L_B[\bn]  - \frac{\rho}{2} (\nabla \bn)^2 \pm \mu e \delta \rho(\br), \qquad \L_B[\bn] = -\frac{\hbar}{2A_M} {\mathcal A}[\bn] \cdot \dot \bn
\label{Lpmn}
\eeq
The first term is a Berry phase term for a spin $1/2$ written in terms of the field of a magnetic monopole $\bA[\bn]$ defined through the relation $\nabla_\bn \times \bA[\bn] = \bn$ \cite{Girvin-LesHouchesNotes}. A simpler way to write this term is by considering its variation
\beq
\delta \L[\bn] = -\frac{\hbar}{2A_M} \bn \cdot (\delta \bn \times \dot \bn)
\eeq
The second term is a gradient term whose coefficient is the pseudospin stiffness. $\delta \rho$ defines the so-called topological charge density given by
\beq
\delta \rho(\br) = \frac{1}{4\pi} \bn \cdot (\partial_x \bn \times \partial_y \bn)
\label{TopDensity}
\eeq
The spatial integral of $\delta \rho$ is an integer. This can be understood by employing the so-called CP${}^1$ representation for $\bn = z^\dagger \bsigma z$, where $z$ is a two component complex unit vector \cite{CPNWitten}. In this representation, the topological density can be written as a total derivative $\delta \rho = \frac{1}{2\pi} \epsilon_{\mu \nu} \partial_\mu (z^\dagger \partial_\nu z)$. The requirement of finite energy means that $\bn$ approaches a constant at infinity which means that the magnitude of both components of $z$ as well as their relative phase approaches a constant. The overall phase of $z$ however can still vary at infinity leading to the general form $z(\br) =_{r \rightarrow \infty} z_0 e^{i \phi(\br)}$. It is then easy to see that the integral of $\delta \rho$ reduces to $\frac{1}{2\pi} \oint d {\vec l} \cdot \nabla \phi$ which is an integer. Another way to understand this integer invariant is by noting that the requirement that $\bn$ approaches a constant at infinity allows us to identify all points at infinity which essentially compactifies the 2D plane to a sphere. The field $\bn$ is then a map from $S^2$ to $S^2$ which is characterized by the homotopy group $\pi_2(S^2) = \mathbb{Z}$. This integer is precisely captured by the integral of the topological density in (\ref{TopDensity}). As discussed at the beginning of Sec.~\ref{sec:QHFM}, a remarkable fact is that the spin texture carries an electric charge which is equal to the topological density times the Chern number (see Ref.~\cite{Girvin-LesHouchesNotes} for details). This explains the last term in the Lagrangian (\ref{Lpmn}) with the topological density coupling to the chemical potential with opposite sign in the opposite Chern sectors. 

To couple the two sectors, we  include the effect of the single-particle dispersion which acts as tunneling between the sectors (cf.~Sec.~\ref{sec:dispersion}) as well as the anisotropy due to deviation from the chiral limit (cf.~Sec.~\ref{sec:devChiral}). The final Lagrangian takes the form
\beq
\label{Lpm}
\L[\bn_+, \bn_-] = \L_+[\bn_+] + \L_-[\bn_-] - J \bn_+ \cdot \bn_- - \lambda (\bn_{+,xy} \cdot \bn_{-,xy} - n_{+,z} n_{-,z})
\eeq

\begin{figure}
    \centering
    \includegraphics[width = 0.5 \textwidth]{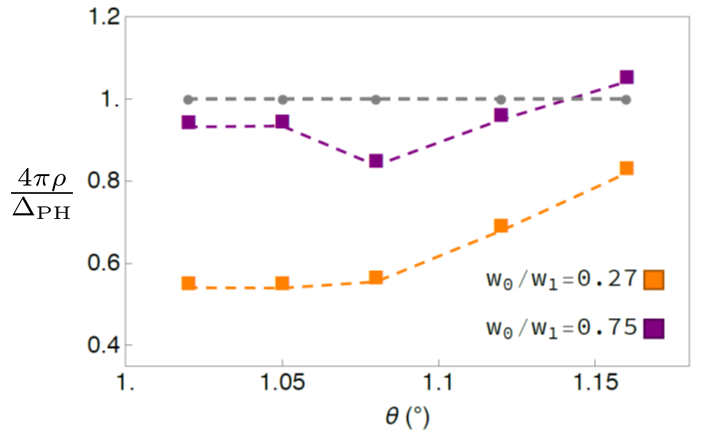}
    \caption{Elastic energy of the skyrmion vs the energy of a particle-hole excitation in the intervalley coherent ground state. Figure adapted from Ref.~\cite{SkPaper}}
    \label{fig:SkPHTBG}
\end{figure}

\subsection{Skyrmion energetics}
To capture the energetics of the skyrmions, we should also add a term $\frac{1}{2} \int d^2 \br V(\br - \br') \delta \rho(\br) \delta \rho(\br')$ for the long range Coulomb interaction. In the absence of coupling between the Chern sectors, the skyrmion energy has two contributions: (i) the elastic contribution coming from the gradient term in the Lagrangian (\ref{Lpmn}) and (ii) the long range Coulomb energy. The former is invariant under the scaling transformation $\bn(\br) \mapsto \bn(\lambda \br)$ which means that it only depends on the skyrmion shape but not its size. The skyrmion configurations which minimizes the elastic energy was first obtained by Polyakov and Belavin \cite{PolyakovBelavin} using the observation that
\beq
 (\partial_a \bn \pm \epsilon_{ab} \bn \times \partial_b \bn)^2 \geq 0, 
\eeq
where we use the convention where latin indices $a, b, \dots$ go over the spatial components $x$, $y$. 
Upon expanding and using the identity $(\ba \times \bb) \cdot (\bc \times \bd) = (\ba \cdot \bc)(\bb \cdot \bd) - (\bb \cdot \bc) (\ba \cdot \bd)$ yields
\beq
(\partial_a \bn)^2 \geq \pm \epsilon_{b c} \bn \cdot (\partial_b \bn \times \partial_c \bn) \quad \implies \quad E_{\rm el} = \frac{\rho}{2} \int d^2 \br (\partial_a \bn)^2 \geq 4\pi \rho |m|
\eeq
 where $m = \frac1{8\pi} \int  \epsilon_{b c} \bn \cdot (\partial_b \bn \times \partial_c \bn)$ is the skyrmion topological charge 
 %(the spatial integral of (\ref{TopDensity}))
 . The equality is only satisfied if $\partial_a \bn = \pm \epsilon_{a b} \bn \times \partial_b \bn$ which was shown by Polyakov and Belavin \cite{PolyakovBelavin} to be equivalent to the condition that the function $w = \frac{n_x + i n_y}{1 - n_z}$ is analytic in $z = x + i y$.
 
 Since the gradient term is scale invariant, the size of the skyrmion is solely determined by the Coulomb repulsion which prefers to make its size as large as possible such that $\delta \rho(\br) \sim O(1/A)$ leading to vanishing Coulomb energy in the thermodynamic limit and a total energy of $4\pi \rho |m|$. The field theory (\ref{Lpmn}) is valid below the gap to electron-hole excitations. In a quantum Hall system, this gap is indeed larger than the skyrmion energy with the ratio $\frac{4\pi \rho}{\Delta_{\rm PH}} = \frac{1}{2}$. For the case of TBG, this ratio depends on a lot of details. In the chiral limit, it turns out to be very close to the quantum Hall value of $1/2$ whereas for the realistic value of $w_0$, it is closer to 1 as shown in Fig.~\ref{fig:SkPHTBG} \cite{SkPaper}. In the following, we will assume that there exists a range of doping such that the skyrmions are the lowest energy charged excitation and take the theory (\ref{Lpmn}) to be a description for the doped system for this doping range. 
 
 Let us now see how coupling the two Chern sectors will change the skyrmion energetics. First, if we switch on the antiferromagnetic coupling $J$, we find that a single skyrmion pays an energy of the order $J R^2$ where $R$ is the skyrmion size \footnote{If we use the Polyakov-Belavin skyrmion, this energy will be logarithmically divergent in the system size. We can however, introduce some regularization which cuts this divergence off at the expense of slightly increasing the elastic energy relative to the minimum bound (see Ref.~\cite{Chatterjee})}. The reason is that at the skyrmion core, the antiferromagnetic condition is violated since the pseudospins are no longer anti-aligned. This contribution acts similar to a Zeeman field for a quantum Hall skyrmion and tries to shrink its size. The competition between this term and the Coulomb repulsion gives the skyrmion a finite size $R \sim (E_C/J)^{1/3}$ (in units of the Moir\'e length) which yields an extra energy contribution $\Delta E \sim J^{1/3} E_c^{2/3}$ on top of the elastic energy.
 
 On the other hand, a charge $2e$ object consisting of a skyrmion in one Chern sector and an antiskyrmion in the opposite sector sitting on top of each other does not pay the energy penalty due to $J$ since the pseudospins are locally anti-aligned. This means that such object has a finite binding energy compared to an infinitely separated skyrmion-antiskyrmion pair which is of the order $\Delta E \sim J^{1/3} E_c^{2/3}$. Thus, the skyrmions and antiskyrmions in opposite Chern sectors pair to form charge $2e$ bound states no matter how small $J$ is, provided that the pseudospin $\SU(2)$ symmetry (which is exact in the chiral limit) is preserved i.e. $\lambda = 0$.
 
 The inclusion of the chiral symmetry breaking $\lambda$ significantly complicates the analysis and it can lead to the deformation of skyrmions into meron pairs or the unbinding of charge $2e$ skyrmions into their individual charge $e$ constituents depending on the detailed values of $\lambda$, $\rho$, and $J$. A detailed quantitative analysis of the skyrmion energetics in this case is beyond the scope of these notes (see Ref.~\cite{SkDMRG} for details).
 
 \subsection{Skyrmion condensation}
 \label{sec:SkCondensation}
 In the following, we will restrict ourselves to the case $\lambda = 0$ realized in the chiral limit where the pseudospin $\SU(2)$ symmetry is exact. Our conclusions will be valid also for $\lambda > 0$ provided it is sufficiently small compared to $J$ as shown in Ref.~\cite{SkDMRG}. In this limit, as discussed above, skyrmions residing in opposite Chern sectors but carrying the same charge $e$ attract forming a bound charge $2e$ object. In addition, we note that the individual charge $e$ skyrmions feel an effective net magnetic field $B_{\rm eff} = \frac{h}{e A_M}$ ($A_M$ is the area of the Moir\'e unit cell) which is opposite in the opposite Chern sectors. For the integer quantum hall effect this must be the case due to gauge invariance: all charge $e$ objects see the same electromagnetic field. However, the band topology of a Chern band is the same as that of a Landau level and so the skyrmion effectively sees the same field. In contrast, the charge $2e$ bound pair does not experience a net magnetic field. Instead, the magnetic field induces a mass of the charge $2e$ object due to a dipole effect as follows: the two charge $e$ skyrmions constituting the charge  $2e$ bound pair carry the same charge but opposite magnetic field which leads to a Lorentz force $F_{\rm Lor} \propto e v B_{\rm eff}$. This force is counterbalanced by a springlike force of attraction which arises from the binding energy $F_{\rm binding} \propto - J x$ \footnote{This arises because the energy penalty associated with separating a skyrmion and antiskyrmion in the opposite Chern sectors by a distance $R$ scales as $R^2$ since it corresponds to the area of the region where the spins are not perfectly anti-aligned}. Equating the two yields the optimal separation as a function of velocity given by $x \propto \frac{e v B_{\rm eff}}{J}$ which yields the energy $E \propto v^2/J$ corresponding to an effective mass $M_{\rm pair} \propto 1/J$.
 
This discussion can be made more precise by considering a pair of skyrmion-antiskyrmion in the opposite Chern sectors whose size and shape are fixed by the energetics with only their position being a free variable, i.e. $\bn_\pm(\br) = \pm \bn_{\rm sk}(\br - \bR_\pm)$ where $\bn_{\rm sk}(\br)$ is some fixed skyrmion configuration. Substituting in the Lagrangian (\ref{Lpm}) yields
\begin{gather}
    L[\bR_+, \bR_-] = \frac{e B_{\rm eff}}{2} [(\bR_+ \times \dot \bR_+) - (\bR_- \times \dot \bR_-)]\cdot \hat z - J F(\bR_+ - \bR_-), \\
    F(\bR) = \int d^2 \br \bn_{\rm sk}(\br) \cdot \bn_{\rm sk}(\br - \bR) 
\end{gather}
Introducing the center of mass coordinate $\bR_s = (\bR_+ + \bR_-)/2$ and the relative coordinate $\bR_d = \bR_+ - \bR_-$, we find that the first term in the Lagrangian becomes $e B_{\rm eff} (\bR_d \times \dot \bR_s) \cdot z$. This means that we can introduce the momentum if the center of mass $\bP_s = \frac{\partial L}{\partial \dot \bR_s} = e B_{\rm eff} \hat z \times \bR_d$. The quantum theory is obtained by promoting $\bR_s$ and $\bP_s$ to operators satisfying $[\hat P_s^i, \hat P_s^j] = [\hat R_s^i, \hat R_s^j] = 0$ and $[\hat R_s^i, \hat P_s^j] = i \hbar \delta_{ij}$ with the Hamiltonian given by 
\beq
\hat H \approx J F\left(\frac{|\hat \bP_s|}{e B_{\rm eff}}\right) = \text{const.} + \frac{J}{2(e B_{\rm eff})^2} F''(0) |\hat \bP_s|^2 + O(|\hat \bP_s|^4), \quad \implies \quad M_{\rm pair} = \frac{(e B_{\rm eff})^2}{J F''(0)}
\eeq
We note that $F''(0) = \int d^2 \br (\nabla \bn)^2$ which yields $F''(0) = 4\pi$ for a skyrmion which minimizes the elastic energy giving $M_{\rm pair} = \frac{(e B_{\rm eff})^2}{4 \pi J}$.

The effective pair mass $M_{\rm pair}$ translates to a superfluid stiffness $\rho_{\rm SC} = \frac{\hbar^2 n}{ M_{\rm pair}}$ where $n$ is the density of charge $2e$ skyrmions taken to be equal to $\nu/(2A_M)$ where $\nu$ is the doping relative to the insulator where superconductivity is developing (for the spinless model, this is taken to be the half-filled insulator). The Kosterlitz-Thouless transition is then given by \cite{NelsonKosterlitz}
\beq
T_{\rm BKT} = \frac{\pi \rho_{\rm SC}}{2k_B} = \frac{\pi \hbar^2 \nu}{2 A_M M_{\rm pair} k_B} = \frac{ \nu J F''(0) A_M}{8 \pi k_B} = \frac{ \nu J A_M}{2 k_B}
\label{TBKT}
\eeq
 
 \subsection{Field theory for skyrmion condensation: CP${}^1$ model}
 \label{sec:FieldTheory}
 To derive a field theoretic description for skyrmion condensation, we consider energies below the scale $J$ where the charge $e$ skyrmions are bound into charge $2e$ objects. In this limit, we can restrict ourselves to pseudospin configurations $\bn = \bn_+ = -\bn_-$ by integrating out the antiferromagnetic fluctuations. This procedure is explained in detail in Refs.~\cite{SkPaper, SkDMRG} which can be summarized here as follows. We start by parametrizing the ferromagnetic fluctuations by writing $\bn_\pm = \pm \bn \sqrt{1 - \bm^2} + \bm$ where $\bm$ is assumed to be small and orthogonal to $\bn$. Expanding the Lagrangian up to quadratic order in $\bm$ yields
 \beq
 \label{Ln}
 \L[\bn] = \frac{\hbar}{2A_M} \bm \cdot (\bn \times \dot \bn) + 2 J \bm^2 + \rho (\nabla \bn)^2 + \frac{\mu e}{2\pi} \bn \cdot (\partial_x \bn \times \partial_y \bn)
 \eeq
 In the first term, we used the fact that the variation of the Berry phase term has a simple form $\delta \L_B[\bn] = \frac{e B_{\rm eff}}{2} \delta \bn \cdot (\bn \times \dot \bn)$. We have also ignored gradients of $\bm$. Integrating out $\bm$ yields \footnote{Note here that since $\bm \cdot (\bn \times \dot \bn)$ vanishes if $\bm$ is parallel to $\bn$, we can relax the condition $\bn \cdot \bm = 0$ and instead integrate over all $\bm$.}
 \beq
 \L[\bn] = \chi (\partial_t \bn)^2 - \rho (\nabla \bn)^2 + \frac{\mu e}{2\pi} \bn \cdot (\partial_x \bn \times \partial_y \bn), \qquad \chi = \frac{\hbar^2}{8J A_M^2}
 \eeq
 
 To study skyrmion condensation, it is useful to rewrite the Lagrangian (\ref{Ln}) in the so-called CP${}^1$ representation. This representation relies on the observation that the sigma model target space is a 2-sphere which is isomorphic to the manifold $\CP^1 = \frac{\SU(2)}{\U(1)}$. It is obtained by introducing a two-component complex unit vector $z$ and writing
 \beq
 \bn = z^\dagger \bsigma z, \qquad z = (z_1, z_2)^T, \qquad z^\dagger z = 1
 \eeq
 This representation has a gauge redundancy corresponding to the overall phase of $z$ since the physical field $\bn$ is unchanged under the gauge transformation $z \mapsto e^{i \phi} z$. The Lagrangian can be written in a simple form by introducing 
 \beq
 \label{az}
 a_\mu = -i z^\dagger \partial_\mu z
 \eeq
 which transforms as a $\U(1)$ gauge field under the aforementioned gauge transformation. We can then write the gradient terms in $\bn$ as $(\partial_\mu \bn)^2 = 4 |(\partial_\mu - i a_\mu) z|^2$. Furthermore, the skyrmion charge turns out to be related to the flux of $a$ via
 \beq
 \delta \rho = \frac{1}{2\pi} \bn \cdot (\partial_x \bn \times \partial_y \bn) = \frac{1}{\pi} \nabla \times a
 \eeq
 To write the final form of the theory, it is convenient to go to imaginary time and introduce the velocity $c = \sqrt{\rho/\chi}$ as well as the dimensionless coupling $g = \frac{\Lambda}{4 \sqrt{\rho \chi}}$ where $\Lambda = 1/\sqrt{A_M}$ can be interpreted as a momentum cutoff, where we have set $\hbar = 1$. In addition, we introduce the rescaled imaginary time variable $r_z = c \tau$ leading to the action
 \beq
 \label{Lza}
 S = \int d^3 \br \L[z, a], \qquad \L[z, a] = \frac{\Lambda}{g} |(\partial_\mu - i a_\mu)z|^2 + \frac{2i e}{2\pi} \epsilon_{\mu \nu \lambda} A_\mu \partial_\nu a_\lambda
 \eeq
 Here, the greek indices go over $x, y, z$. Note that the action is dimensionless since $\hbar$ was set to 1. Here, we have generalized the term coupling to the chemical potential $\mu$ to a general coupling to a background gauge field $A_\mu$ with $A_0 \propto \mu$.
 
 An important observation here is the following. Although the gauge field was introduced in such a way that it is tied to the $z$ field through Eq.~\eqref{az}, we can take it to be an independent field \cite{CPNWitten}. The reason is that the constraint (\ref{az}) is anyway reproduced via the equations of motion obtained from the Lagrangian (\ref{Lza}) which take the form
 \beq
 \frac{2\Lambda}{g}(a_\mu + i z^\dagger \partial_\mu z) + \frac{2i e}{2\pi} \epsilon_{\mu \nu \lambda} \partial_\nu A_\lambda = 0
 \eeq
 which reproduces (\ref{az}) for a time-independent and spatially constant background field $A_\mu$. Another way to see this is to note that the flux of $a_\mu$ has to be accompanied with the winding of $z$ field to obtain a finite energy configuration which imposes the constraint (\ref{az}) at long distance. 
 
 \subsubsection{Phases of the CP${}^1$ model}
 
 The main advantage of the CP${}^1$ theory is that it maps the non-linear sigma model (\ref{Ln}) to a theory of complex bosons coupled to a gauge field which is much easier to analyze. Generally, such a theory admits two phases: (i) an ordered phase where $z$ has a finite expectation values which Higgses the $\U(1)$ gauge field $a$ and (ii) a disordered phase where $z$ is gapped 
 \footnote{Here we say that $z$ is gapped in a ground state $\ket{\Omega}$ if the state $z^\dag (\bq)\ket{\Omega}$ costs energy $\geq \Delta$ for any wavenumber $\bq$, for some $\Delta > 0$. In other words, the particles created by the creation operator $z^\dag$ have a minimum nonzero energy. Goldstone modes are not gapped because they cost zero energy for $\bq \to 0$.}
 and $a$ is in the Coulomb phase. \footnote{By this we mean that $a$ is not Higgsed such that the photons of $a$ are massless. These photons then mediate a long range Coulomb force between particles charged under $a$.}
 The ordered phase (i) is an insulator which can be seen by integrating out the gapped field $a$ yielding a Maxwell term $\sim \frac{e^2 g}{|\langle z \rangle|^2 \Lambda} (\epsilon_{\mu \nu \lambda} \partial_\nu A_\lambda)^2$ which is describes an insulating dielectric. The pseudospin order parameter has a finite expectation value given by $\langle \bn \rangle = \langle z \rangle^\dagger \bsigma \langle z \rangle$. The disordered phase (ii) is a superconductor. This can be seen by first integrating out the gapped variable $z$ which gives a Maxwell term for $a$ \cite{CPNdAdda, CPNWitten, PolyakovBook}. The resulting Lagrangian has the form
 \beq
 \L = \frac{1}{2\kappa} (\epsilon_{\mu \nu \lambda} \partial_\nu a_\lambda)^2 + \frac{i e}{\pi} \epsilon_{\mu \nu \lambda} A_\mu \partial_\nu a_\lambda
 \label{MaxwellCS}
 \eeq
 which is dual to a theory of complex boson coupled to the background field with charge $2e$ i.e. a superconductor. This is established by defining the current $J_\mu = \frac{1}{2\pi} \epsilon_{\mu \nu \lambda} \partial_\nu a_\lambda$ which is manifestly conserved i.e. $\partial_\mu J_\mu=0$. The current conservation can be enforced in the field theory by including the integral representation of the delta function implemented by adding the term $i \phi \partial_\mu J_\mu$ to the Lagrangian where the field $\phi$ is an integration variable. Integrating out $J_\mu$ yields $\L = \frac{\kappa}{8\pi^2} (\partial_\mu \phi - 2 e A_\mu)^2$ which describes a superconductor with superfluid stiffness $\rho_{\rm SC} = \frac{\kappa c}{4 \pi^2}$.
 
 Let us note the following. Although the $z$ variables are gapped inside the superconducting phases, $\langle z \rangle = 0$, this does not necessarily imply the absence of pseudospin order which is described instead by the order parameter $\langle \bn \rangle = \langle z^\dagger \bsigma z \rangle$. In fact, we do not expect the skyrmion condensation to destroy the pseudospin order since a skyrmion with finite size only alters the spin in a finite region of space. Within the CP${}^1$ theory, this can be understood as follows. The skyrmion creation operator can be identified with the operator which creates a $2\pi$ magnetic flux of the $a$ field; such an operator is called a monopole operator. To obtained a finite energy excitation, such $2\pi$ flux has to be associated with a vortex in both $z_1$ and $z_2$ where the phase of each rotates by $2\pi$ at infinity. The proliferation of such vortices disorders the phases of both $z_1$ and $z_2$ leading to $\langle z_{1,2} \rangle = 0$, but not the phase of $z_i^* z_j$ for $i,j = 1,2$. Thus, the superconductor obtained by skyrmion condensation is compatible with a finite pseudospin order $\langle \bn \rangle \neq 0$.
 
 \subsubsection{Large $N$ analysis}
  
 In order to achieve a more quantitative understanding of the phase diagram, we will employ a large $N$ approximation by promoting the CP${}^1$ variable $z = (z_1, z_2)$ to a CP${}^{N-1}$ variable $z = (z_1, \dots, z_N)$ satisfying $z^\dagger z = N-1$. We note that, as pointed out in the previous section, we can take the gauge field $a_\mu$ to be independent of $z$. In addition, we can enforce the constraint by using the integral representation of the delta function by adding $i \kappa (z^\dagger z - (N-1))$ to the Lagrangian leading to
 \beq
 \L = \frac{\Lambda}{g}  \left\{ \sum_{l=1}^N z_l^* (-\D^2 + \Delta^2) z_l - \Delta^2 (N-1) \right\}, \qquad \Delta^2 = i \kappa \frac{g}{\Lambda}, \qquad \D_\mu = \partial_\mu - i a_\mu
 \eeq
We then proceed to integrate out the variables $z_2$ to $z_N$ and rescaling the variable $z_1$ by $\sqrt{N-1}$ leading to the Lagrangian
\beq
S = (N-1)  \left\{\tr \ln (-\D^2 + \Delta^2) + \frac{\Lambda}{g} \int d^3 \br [z_1^* (-\D^2 + \Delta^2) z_1 - \Delta^2] \right\}
\label{SN}
\eeq
which enables us to evaluate the path integral using the saddle point approximation. The saddle point equations relative to $\Delta$ and $z_1$ take the form \footnote{there is also a saddle point equation relative to $a_\mu$ that does not play a significant role in our analysis so will be omitted}
\beq
(-\D^2 + \Delta^2) z_1 = 0, \qquad \frac{\Lambda}{g}(|z_1|^2 - 1) = G_0(\Delta) = \tr \frac{1}{-\D^2 + \Delta^2}
\label{Saddle}
\eeq
We can first consider the case when the chemical potential is sufficiently small (compared to the gap to adding charge $2e$ skyrmions) such that there is no skyrmions in the system. Since the skyrmions are associated with the flux of $a$, we can choose $a = 0$  leading to the simplification \cite{CPNdAdda, CPNWitten, PolyakovBook}
\beq
(-\nabla^2 + \Delta^2) z_1 = 0, \qquad \frac{\Lambda}{g}(1 - |z_1|^2) = \int \frac{d^3 \bk}{(2\pi)^3} \left[\frac{1}{\bk^3 + \Delta^3} - \frac{1}{\bk^3 + \Lambda^3}\right] =  \frac{1}{4\pi} (\Lambda - |\Delta|)
\eeq
Here, we have introduced a Pauli-Villars regularizer for the momentum integral in $G_0$. There are two possible solutions to the first equation: (i) $z_1 = 0$ or (ii) $\Delta = 0$, $z_1$ is a constant. The first solution leads to $|\Delta| = \Lambda (1 - 4\pi/g)$ which can only be realized for $g \geq 4\pi$, whereas the second solution leads to $|z_1| = 1 - g/4\pi$ which can only be realized for $g \leq 4\pi$ since $|z_1|\leq 1$. Thus, the theory has two phases depending on the value of $g$. For $g < 4\pi$, the spin is ordered  since $|z_1| \neq 0$ which Higgses $a$ and leads to an insulating phase. For $g > 4\pi$, the $z$ variables are gapped and we can expand the action (\ref{SN}) to the next order in the field $a$ leading to a Maxwell term \cite{CPNdAdda, CPNWitten, PolyakovBook} as discussed earlier (cf.~Eq.~\ref{MaxwellCS}) with $\kappa = \frac{24\pi |\Delta|}{N}$, leading to the superfluid stiffness $\rho_{\rm SC} = \frac{\kappa c}{4\pi^2} = \frac{6|\Delta| c}{\pi N}$.

To see what happens when skyrmions are introduced into the system via doping, we consider saddle points characterized by a finite and constant value of $\nabla \times a$ which is related to the filling $\nu$ via $\nabla \times a = \frac{\nu \pi}{A_M}$. The saddle point equations have the same form as in Eq.~\ref{Saddle} with $G_0(\Delta)$ given by \cite{SkPaper}
\beq
G_0(\Delta) = \frac{\Lambda}{4} \sqrt{\frac{\nu}{2\pi}} \left\{ \zeta \left(\frac{1}{2}, \frac{1}{2} + \frac{\Delta^2}{2\pi \nu \Lambda^2} \right) - \zeta \left(\frac{1}{2}, \frac{1}{2} + \frac{1}{2\pi \nu} \right) \right\}
\eeq
where $\zeta$ is the Hurwitz zeta function. At small doping $\nu \ll 1$, we can solve this equation perturbatively to get \cite{SkPaper}
\beq
\Delta^2 = \Lambda^2 \left[ -\pi \nu + \frac{1}{16} g^2 \nu^2 + O(\nu^3) \right]
\label{Deltanu}
\eeq
which is valid for $g < 4\pi$. Working in the small doping limit, we can again expand the saddle point action in deviations of $a$ from its background value to get a Maxwell term as in Eq.~\ref{MaxwellCS} with $\kappa = \frac{24 \pi \sqrt{\Delta^2 + \pi \nu \Lambda^2}}{N}$ which upon substituting in (\ref{Deltanu}) yields
\beq
\rho_{\rm SC} = \frac{\kappa c}{4\pi^2} = \frac{3 g c \nu \Lambda}{2\pi N} = \frac{3 \Lambda^2 \nu}{8\pi \chi N} = \frac{3 J A_M \nu}{\pi N} ,\quad \implies \quad T_{\rm BKT} = \frac{3 J A_M \nu}{2 k_B N}
\eeq
This expression for $T_{\rm BKT}$ has the same dependence on $J$ and $\nu$ as the one obtained from the skyrmion effective mass in Eq.~\eqref{TBKT}. If we substitute with $N = 2$, we get $T_{\rm BKT} = \frac{3 J A_M \nu}{4 k_B}$ which is larger by a factor of $3/2$. This is not surprising since a large $N$ calculation is not expected to be quantitatively accurate for $N = 2$.

\subsubsection{Numerical evidence}
Compelling numerical evidence for a skyrmion mechanism for superconductivity in a model of antiferromagnetically coupled quantum Hall ferromagnets was recently obtained in a numerical DMRG study by Chatterjee {\it et. al.} \cite{SkDMRG}. In this work the  Chern bands of TBG are approximated by lowest Landau levels in opposite magnetic fields. This can be partially motivated by the discussion of Sec.~\ref{Sec:SingleParticle} which showed that the chiral model of TBG shares not only the topology of the lowest Landau level but also its band geometry.

The main result of this work can be summarized as follows. %Fig.~\ref{fig:DMRG} which shows 
First, the nature of the charged excitations in the insulator were determined as a function of the antiferromagnetic coupling $J$ and the $\SU(2)$-breaking anisotropy parameter $\lambda$ where $\lambda > 0$ ($\lambda < 0$) denote easy-plane (easy-axis) anisotropy. As discussed earlier, chiral TBG corresponds to $\lambda = 0$ while the realistic model corresponds to the easy-plane case $\lambda > 0$. Both field theoretical calculations of topological textures and DMRG numerics were utilized to find the range of $J,\,\lambda$ where charge 2e skyrmions are the lowest energy charge excitations. This is one indicator of superconductivity . The other, more direct indicator is from the direct measurement of pair correlation functions at small but finite doping. The two were found to be broadly consistent with each other and give the following phase boundaries for  superconductivity. 
 The main observation is that in the $\SU(2)$ limit, superconductivity is stablized no matter how small $J$ is (in units of the Coulomb scale) consistent with the analysis of Secs.~\ref{sec:SkCondensation} and \ref{sec:FieldTheory}. Furthermore, superconductivity remains robust for easy plane anisotropy for $\lambda < \lambda_c \sim J$ but is  fragile in the opposite limit, i.e. in the presence of easy-axis anisotropy. Finally, an independent computation of the binding energy of the charge $2e$ skyrmions from the non-linear sigma model shows that the region in the phase diagram where the $2e$ bound state is stable roughly coincides with the region where superconductivity is stabilized at finite doping, thereby providing strong evidence for its skyrmion origin. It should be noted however that here  the  Berry curvature inhomogeneity and deviations from the ideal band geometry are neglected, which can have important implications for the physics of TBG \cite{SoftModes}. Nevertheless, this numerical study shows that the skyrmion mechanism is a viable and internally consistent theory of superconductivity, and magic angle graphene does incorporate its key requirements. We also point the reader to two other works that may be of interest in this context \cite{Christos2020,Liu_2019}. 

%\begin{figure}
%    \centering
%    \includegraphics[width = 0.7 \textwidth]{DMRGFig.png}
%    \caption{(a) Pair binding energy of the charge $2e$ skyrmions computed from the non-linear sigma model and (b) superconducting pairing in DMRG for different values of the cylinder circumference as a function of $J$ and $\lambda$ \cite{SkDMRG}.}
%    \label{fig:DMRG}
%\end{figure}

\section{Multilayer generalizations: alternating twist multilayer graphene}
\begin{shaded}
In this section, we discuss a multilayer generalization of TBG with an {\em alternating} twist,  introduced in Ref.~\cite{Khalaf_multilayer}. These multilayer systems turned out to have very similar properties to twisted bilayer graphene both in the chiral and non-chiral limits and retains most of its important symmetries, particularly $C_2 \T$.
\end{shaded}
The alternating twist multilayer graphene (ATMG) is defined by stacking $n$ graphene layers such that the relative twist angle between layers have equal magnitude but alternating signs i.e. $\theta_l = (-1)^l \theta/2$, $l = 1,\dots, n$. In general, this generates $n-1$ Moir\'e patterns which have the same wavelength but may be displaced relative to each other. We restrict ourselves to the case where such relative displacements vanish causing these Moir\'e patterns to align. This configuration was shown in Ref.~\cite{Carr2020} to be the energetically most stable configuration for $n=3$ and this conclusion is likely to hold for general $n$. In the following, we will always assume the case where the Moir\'e patterns are all aligned.

The Hamiltonian for ATMG with $n$ layers for a single flavor (single spin-valley)  is given by
\beq
H(\br) = \left(\begin{array}{ccccc}
-i v_F \bsigma_+ \cdot \nabla  & T(\br) & 0 & 0 & \dots \\
T^\dagger(\br) & -i v_F \bsigma_- \cdot \nabla & T^\dagger(\br) & 0 & \dots \\
0 & T(\br) & -i v_F \bsigma_+ \cdot \nabla  & T(\br) \\
0 & 0 & T^\dagger(\br) & -i v_F \bsigma_- \cdot \nabla & \dots \\
\dots & \dots & \dots & \dots & \dots
\end{array} \right)
\label{HATMG}
\eeq
where $\bsigma_\pm = e^{\pm \frac{i}{4} \theta \sigma_Z} \bsigma e^{\mp \frac{i}{4} \theta \sigma_z}$ and $T(\br)$ is given by
\begin{gather}
    T(\br) = \left(\begin{array}{cc}
w_0 U_0(\br) & w_1 U_1(\br) \\
w_1 U^*_1(-\br) & w_0 U_0(\br)
\end{array} \right), \qquad U_m(\br) = \sum_{n=1}^3 e^{\frac{2 \pi i}{3} m (n - 1)} e^{-i \bq_n \cdot \br}, \\ \bq_n = 2 k_D \sin \left(\frac{\theta}{2}\right) R_{\frac{2 \pi (n-1)}{3}}(0,-1)
\end{gather}
where $R_\phi$ denotes counter clockwise rotation by $\phi$ and  $k_D = \frac{4\pi}{3 \sqrt{3} a_{\rm CC}}$.

A remarkable result proved in Ref.~\cite{Khalaf_multilayer} is that the Hamiltonian (\ref{HATMG}) maps to $m$ copies of TBG with rescaled tunneling matrices $T$ for $n = 2m$ and to $m$ copies of TBG (also with rescaled tunneling matrices $T$) in addition to a single Dirac cone for $n = 2m + 1$. To establish this mapping, we follow Ref.~\cite{Khalaf_multilayer} and perform the transformation $\tilde H = P^T H P$ with $P$ defined as
\beq
P = \sum_{k=1}^{\lfloor n/2 \rfloor} \delta_{2k, \lceil n/2 \rceil + k} + \sum_{k=1}^{\lceil n/2 \rceil} \delta_{2k-1, k}
\eeq
This transformation simply relabels the layers such that the Hamiltonian has a block matrix structure in the even-odd layer index
\beq
\tilde H = P^T H P = \left( \begin{array}{cc} -i v_F \bsigma_+ \cdot \nabla & T(\br) \otimes W \\
T^\dagger(\br) \otimes W^\dagger & -i v_F \bsigma_- \cdot \nabla \end{array} \right), 
\eeq
with $W$ being an $\lceil n/2 \rceil \times \lfloor n/2 \rfloor$ describing the structure of the tunneling between layers. Since tunneling only takes place between nearest neighboring layers, $W$ has the simple form $W_{ij} = \delta_{i,j} + \delta_{i,j+1}$. The mapping can be established by writing the singular value decomposition for $W = A \Lambda B^\dagger$ which implies that
\beq
H' = \left( \begin{array}{cc} A^\dagger & 0 \\ 0 & B^\dagger \end{array} \right) \tilde H \left( \begin{array}{cc} A & 0 \\ 0 & B \end{array} \right) = \left( \begin{array}{cc} -i v_F \bsigma_+ \cdot \nabla & T(\br) \otimes \Lambda \\
T^\dagger(\br) \otimes \Lambda^\dagger & -i v_F \bsigma_- \cdot \nabla \end{array} \right)
\eeq
Here, $\Lambda$ is a diagonal matrix with eigenvalues $\lambda_k = 2 \cos \frac{\pi k}{n+1}$ where $k = 1, \dots, \lfloor n/2 \rfloor$. For even $n$, this means that the Hamiltonian $H'$ decomposes into a sum of $n/2$ TBG Hamiltonians with the tunneling $T(\br)$ rescaled by $\lambda_k$. For odd $n$, the matrix $\Lambda$ is a rectangular $(n-1)/2 \times (n+1)/2$ matrix. As a result, the Hamiltonian $H'$ reduces to $(n-1)/2$ TBG Hamiltonians with the tunneling $T(\br)$ rescaled by $\lambda_k$ in addition to a single Dirac cone that is decoupled from the rest of the system. The general unitary matrix which maps $H$ to a decoupled sum of TBGs possibly in addition to a Dirac cone has the explicit form 
\beq
H_{\rm dec} = V^\dagger H V, \qquad V = P \left(\begin{array}{cc}
A & 0 \\ 0 & B \end{array} \right) P^T
\eeq
The different TBG systems can be approximately understood as corresponding to different momenta along the $z$ direction with the subtlety that there is no true translation in the vertical direction due to the absence of periodic boundary condition (since the topmost and bottom layers are not connected by tunneling).

\subsection{Trilayer $n=3$}
The simplest example of ATMG corresponds to the trilayer case with $n = 3$. The Hamiltonian for this case is given explicitly by
\beq
H(\br) = \left(\begin{array}{ccc}
-i v_F \bsigma_+ \cdot \nabla  & T(\br) & 0 \\
T^\dagger(\br) & -i v_F \bsigma_- \cdot \nabla & T^\dagger(\br) \\
0 & T(\br) & -i v_F \bsigma_+ \cdot \nabla \end{array} \right)
\label{HK}
\eeq
The matrix $P$ given explicitly by
\beq
P = \left(\begin{array}{ccc}
1 & 0 & 0 \\ 0 & 0 & 1 \\ 0 & 1 & 0 \end{array} \right) \, \implies \, \tilde H = P^T H P = \left(\begin{array}{ccc}
-i v_F \bsigma_+ \cdot \nabla  & 0 & T(\br) \\
0 & -i v_F \bsigma_+ \cdot \nabla & T(\br) \\
T^\dagger(\br) & T^\dagger(\br) & -i v_F \bsigma_- \cdot \nabla \end{array} \right)
\eeq
The tunneling matrix $W$ and its singular value decomposition $W = A \Lambda B^\dagger$ are given by
\beq
W = \left(\begin{array}{c}
1 \\ 1 \end{array} \right), \, \implies \, A = \frac{1}{\sqrt{2}}\left(\begin{array}{cc}
1 & -1 \\ 1 & 1 \end{array} \right), \, \Lambda = \left(\begin{array}{c}
\sqrt{2} \\ 0 \end{array} \right), \, B = 1
\eeq
Combining these two, we can construct the unitary matrix $V$ which decouples $H$ given explicitly by
\begin{gather}
    V = P \left(\begin{array}{cc}
A & 0 \\ 0 & B \end{array} \right) P^T = \left(\begin{array}{ccc}
\frac{1}{\sqrt{2}} & 0 & -\frac{1}{\sqrt{2}}  \\
0 & 1 & 0 \\
\frac{1}{\sqrt{2}} & 0 & \frac{1}{\sqrt{2}} \end{array} \right) \\ \quad \implies \quad H_{\rm dec} =  V^\dagger H V = \left(\begin{array}{ccc}
-i v_F \bsigma_+ \cdot \nabla  & \sqrt{2}  T(\br) & 0 \\
\sqrt{2} T^\dagger(\br) & -i v_F \bsigma_- \cdot \nabla & 0 \\ 
0 & 0 & -i v_F \bsigma_+ \cdot \nabla \end{array} \right)
\end{gather}
The mapping can be simply understood in terms by noting that the TBG blocks consists of the middle layer and the bonding combination of the top and bottom layers whereas the single Dirac cone consists of the anti-bonding combination. The rescaling of the interlayer tunneling by $\sqrt{2}$ means that the magic angles of the trilayer system are larger by a factor of $\sqrt{2}$ compared to TBG magic angles. In particular, the first magic angle is $\theta \approx 1.5^o$.

\subsection{Tetralayer $n=4$}
We can similarly work out the Hamiltonian for the tetralayer case $n = 4$. The Hamiltonian for this case is given explicitly by
\beq
H(\br) = \left(\begin{array}{cccc}
-i v_F \bsigma_+ \cdot \nabla  & T(\br) & 0 & 0\\
T^\dagger(\br) & -i v_F \bsigma_- \cdot \nabla & T^\dagger(\br) & 0 \\
0 & T(\br) & -i v_F \bsigma_+ \cdot \nabla & T(\br) \\
0 & 0 & T^\dagger(\br) & -i v_F \bsigma_- \cdot \nabla 
\end{array} \right)
\label{HK}
\eeq
The matrix $P$ given explicitly by
\beq
P = \left(\begin{array}{cccc}
1 & 0 & 0 & 0 \\ 0 & 0 & 1 & 0 \\ 0 & 1 & 0 & 0 \\ 0 & 0 & 0 & 1 \end{array} \right)
\eeq
The tunneling matrix $W$ and its singular value decomposition $W = A \Lambda B^\dagger$ are given by
\beq
W = \left(\begin{array}{cc}
1 & 0 \\ 1 & 1 \end{array} \right) \, \implies \, A = \left(\begin{array}{cc}
\sqrt{\frac{3 - \varphi}{5}} & -\sqrt{\frac{2 + \varphi}{5}} \\ \sqrt{\frac{2 + \varphi}{5}} & \sqrt{\frac{3 - \varphi}{5}} \end{array} \right), \, \Lambda = \left(\begin{array}{cc}
\varphi & 0 \\ 0 & 1/\varphi \end{array} \right), \, B = \left(\begin{array}{cc}
\sqrt{\frac{2 + \varphi}{5}} & -\sqrt{\frac{3 - \varphi}{5}}  \\ \sqrt{\frac{3 - \varphi}{5}} & \sqrt{\frac{2 + \varphi}{5}} \end{array} \right)
\eeq
where $\varphi$ is the golden ratio $\varphi = \frac{1 + \sqrt{5}}{2}$. The decoupling matrix $V$ is
\begin{gather}
    V = \left(\begin{array}{cccc}
\sqrt{\frac{3 - \varphi}{5}} & 0 & -\sqrt{\frac{2 + \varphi}{5}} & 0 \\
0 & \sqrt{\frac{2 +  \varphi}{5}} & 0 & -\sqrt{\frac{3 - \varphi}{5}}\\
\sqrt{\frac{2 + \varphi}{5}} & 0 & \sqrt{\frac{3 - \varphi}{5}} & 0 \\
0 & \sqrt{\frac{3 - \varphi}{5}} & 0 & \sqrt{\frac{2 + \varphi}{5}}\end{array} \right) \\ \quad \implies \quad H_{\rm dec} =  V^\dagger H V = \left(\begin{array}{cccc}
    -i v_F \bsigma_+ \cdot \nabla & \varphi T(\br) & 0 & 0 \\
\varphi T^\dagger(\br) & -i v_F \bsigma_- \cdot \nabla & 0 & 0 \\
0 & 0 & -i v_F \bsigma_+ \cdot \nabla & \frac{1}{\varphi} T(\br) \\
0 & 0 & \frac{1}{\varphi} T^\dagger(\br) & -i v_F \bsigma_- \cdot \nabla 
\end{array} \right)
\end{gather}
Thus, the tetralayer system maps to a pair of TBGs whose interlayer tunneling is scaled by $\varphi$ and $1/\varphi$. This yields two sequences of magic angles obtained from TBG magic angles either by multiplying or dividing by the golden ratio $\varphi$ \cite{Khalaf_multilayer}. This mapping can also be understood in terms of bonding and antibonding orbitals as follows. The first TBG (with $T$ rescaled by $\varphi$) is built using the bonding orbitals between the first and third (layer 1) and second and fourth (layer 2) whereas the second TBG (with $T$ rescaled by $1/\varphi$) is built using the antibonding orbitals between the first and third (as layer 1) and second and fourth (as layer 2).

% This term is invariant under spinful time-reversal symmetry and thus has the same sign for both Chern sector. This can be seen by considering its variation given by
% \beq
% \delta \L_B = \frac{i \hbar}{2} \bn \cdot (\delta \bn \times \dot \bn)
% \eeq
% Under spinful time-reversal $\bn \mapsto -\bn$ and $\partial_t$  (this is the Kramers time-reversal we considered earlier
% arising from the motion of spin in a background magnetic field. $\mathcal A$ describes the field of a magnetic monopole This term only relies on topology and 
% Apart Thus, a field theoretic description of the system is closely related to that of a pair of quantum Hall ferromagnets in opposite magnetic field. In fact, the descri

%\subsection{Skyrmion pairing}

%\subsection{Field theory for the superconductor: large $N$ theory}

%{\bf Todo 4:} Discuss calculation of form factors and of  gaps and describe insulator in this limit? This model has SU(4) symmetry, if kinetic terms are not neglected. Specialize to the simpler spinless model and justify the anisotropies.  

%{\bf Todo5:} Justify Skyrmion energetics introduce Cp1 representation. 

\section{Discussion and Conclusions} 
%\AV{briefly summarize Mike's DMRG?} 
In his 1977 Nobel Lecture, Phil Anderson observes:\begin{quote}
The art of model-building is the exclusion of real but irrelevant parts of the problem, and entails hazards for the builder and the reader.
%    \em One of my strongest stylistic prejudices in science is that many of the facts
%Nature confronts us with are so implausible … that the mere
%demonstration of a reasonable mechanism leaves no doubt of the correct
%explanation.
\end{quote}{}
In keeping with this spirit, we note that our perspective on magic angle graphene   effectively begins by approaching the problem from the chiral limit, although several realistic features are subsequently included. In this limit, the flat bands take the form of zero modes of a Dirac operator, which can be analytically obtained. Further,  not just the single particle states but even the interacting ground state at certain integer fillings can be reliably found in this limit, which are insulators with a generalized flavor order. The topology of the bands ensures the flavor order and the charge physics are inter-related.  This has implications for finite doping, and we discussed a plausible topological mechanism for superconductivity. Other states at fractional filling, notably fractional Chern insulators, are also potentially stabilized due to the special band geometry of the flat bands that resemble that of the lowest Landau level. We believe this viewpoint broadly captures the observed physics and makes predictions for future experiment.   At the same time, let us mention some of the complexities that have been left out. First, 
 although tunneling between opposite Chern bands is retained, which is a significant source of dispersion,  the intra-Chern band dispersion is neglected. This contribution may be particularly important away from charge neutrality where the Hartree correction leads to substantial dispersion of the bands \cite{XieMacdonald, GuineaHF}. Second, the potentially important role of moire lattice strain, as well as of phonons was not  discussed. These and other approximations (such as anisotropies in the sigma model) could alter the energetics of charge carriers, i.e. the competition between charge 2e skyrmions and electrons. For small doping, charge could be added as electrons/holes, and only later intersect the 2e-skyrmion energy. Developing a theory of such Fermi-Bose models will be an interesting and important task for the future, related to the quest for a weak coupling version of the same pairing mechanism.  

Another important direction is to design experimental probes to identify  subtle flavor orders (such as intervally coherence, in particular K-IVC, featuring spatially modulated currents) which would be instrumental in both detecting such order in insulators, and tracking their  evolution on doping. One route to identifying the flavor ordered state is by detecting the collective modes associated with the ground state \cite{SoftModes}. These modes can serve as a fingerprint of the flavor order and also help pin down key parameters like the flavor stiffness, and anisotropies, $J$ and $\lambda$. 

%Finally, keeping the broader context in mind, the question that we would really like to answer is - what are the essential ingredients leading to superconductivity in MATBG and can we duplicate these at higher energy scales? Anderson who in his 1977 Nobel lecture describes his own scientific style as

%Indeed a worthy  goal is to isolate a reasonable mechanism for superconductivity in MATBG and its key ingredients, and  compare and contrasts with other mechanisms, for example those proposed in the context of the cuprate high temperature superconductors.

%no doubt this explains the profusion of "Anderson models" and their big role in clarifying fundamental phenomena, although this prescription may not work as well for others. 
\bigskip

\newpage
\section{Acknowledgements} We would like to  thank  N. Bultinck, S. Chatterjee, A. Kruchkov, J. Y. Lee, S. Liu, G. Tarnopolsky and M. Zaletel for collaborations summarized in this work. AV was supported by a Simons Investigator award and by the Simons Collaboration on Ultra-Quantum Matter, which is a grant from the Simons Foundation (651440,  AV). EK  was supported by the German National Academy of Sciences
Leopoldina through grant LPDS 2018-02 Leopoldina fellowship.  P.J.L. was partly supported by the Department of Defense
(DoD) through the National Defense Science
and Engineering Graduate Fellowship (NDSEG)
Program.
\appendix

\section{Symmetries of the chiral model}
\label{Sec:AppendixSymmetries}
The main feature of the chiral TBG Hamiltonian is the presence of a chiral or sublattice symmetry which anticommutes with the single-particle Hamiltonian $\{ \sigma_z, H \} = 0$. In addition, weak-spin orbit coupling in graphene implies that $H$ is invariant under ${\rm SU}(2)$ spin rotations, i.e. $H \propto s_0$. The large momentum separation between the graphene valleys leads to valley decoupling at low energies relevant to TBG physics which implies independent spin and charge conservation in each valley i.e. $H$ is diagonal in the valley index $\tau_z$. The two valley are related by spinless time-reversal symmetry which is implemented as $\T = \tau_x \K$ with $\K$ denoting complex conjugation, such that $H_{K'}(\bk) = H_K(-\bk)^*$. The single particle Hamiltonian is also invariant under two-fold rotation symmetry which flips sublattice and valley $C_2 = \sigma_x \tau_x$. The combination of the two $C_2 \T = \sigma_x \K$ leaves the valley index invariant and relates the two sublattice wavefunctions at the same momentum $\bk$.

A more subtle symmetry of the chiral TBG Hamiltonian is a particle-hole symmetry $H(-\bk)^* = \mu_y \sigma_x H(\bk) \mu_y \sigma_x$ where $\mu$ denote the Pauli matrices in layer space \cite{Song, Hejazi}. On the level of the single-particle Hamiltonian this is an antiunitary anticommuting symmetry given by $\P = i\mu_y \sigma_x \K$ whereas in second quantized language, this is a unitary particle-hole symmetry acting as
\beq
\P^{-1} f_{\bk} \P = \mu_y \sigma_x f^\dagger_{-\bk}, \qquad  \P^{-1} i \P = i
\eeq
where $f_\bk$ denotes the annihilation operators for microscopic electrons. In addition to the sublattice, valley, and spin indices, these electrons also carry a layer index.% (\EK{and a reciprocal lattice index. I do not know if we need to mention it}).

Since the chiral symmetry $\sigma_z$ is anticommuting and unitary ( anti-unitary particle-hole in second quantization), $\T$ is commuting and anti-unitary (anti-unitary in second quantization), and $\P$ is anti-commuting and anti-unitary (unitary particle-hole in second quantization), their product $R = i\sigma_z \P \T = \tau_x \sigma_y \mu_y$ is a $Z_2$ unitary commuting symmetry. This extra ``hidden" symmetry relates different valley, sublattice and layer indices and leads to enhanced symmetry of the interacting problem as we will see below.

%\EK{The following discussion is a bit subtle. It is related to the fact that the microscopic implementation of the symmetries differs from the representation in the projected basis. The latter depends on the gauge choice and has some subtleties due to band topology. It is necessary to understand the form of $h$ and $\Lambda$, but I do not know if you want to include it}

\subsection{Symmetries and the Form Factor}
\label{Sec:twoparticleSymmetries}
Let us now see how that symmetries restrict the structure of the form factor $\Lambda_\bq(\bk)$. Due to the $\SU(2)$ spin rotation symmetry, the spin index does not play any role in the following discussion so it can be dropped. This essentially corresponds to considering a spinless version of the problem. In the following, we will write out the flat band index $\alpha$ in terms of the sublattice-valley indices $\sigma$ and $\tau$. We note that sublattice symmetry implies that $\Lambda_\bq(\bk)$ is diagonal in the sublattice index since 
\beq
\langle u_{A, \tau, \bk} | u_{B, \tau, \bk'} \rangle = \langle u_{A, \tau, \bk} | \sigma_z^2 | u_{B, \tau, \bk'} \rangle = - \langle u_{A, \tau, \bk} | u_{B, \tau, \bk'} \rangle = 0
\eeq
where we used $\sigma_z | u_{A/B, \tau, \bk'} \rangle = \pm | u_{A/B, \tau, \bk'} \rangle$. In addition, valley symmetry implies that $\Lambda_\bq(\bk)$ is also diagonal in valley. We now note that since $R$ symmetry flips both valley and sublattice, we have
\beq
R u_{\sigma, \tau, \bk} = e^{i \phi_{\sigma, \tau, \bk}} u_{-\sigma, -\tau, \bk} 
\label{RFB}
\eeq
for some phase $\phi_{\sigma,\tau,\bk}$. Since we are free to choose the gauge independently in the two valleys $\tau = \pm = K/K'$ and sublattices $\sigma = \pm = A/B$, we can set $\phi_{\sigma, \tau, \bk}$ to 0 by fixing the gauge e.g. in valley $K'$ relative to $K$. This means that we can choose the action of $R$ on the flat bands to be $\sigma_x \tau_x$ \footnote{Note that this gauge choise is slightly different from the one used in Ref.~\cite{KIVCpaper}}. Similarly, we can deduce the representation of $C_2 \T$ on the flat bands by noting that it flips sublattice but not momentum
\beq
C_2 \T u_{\sigma, \tau, \bk} = e^{i \eta_{\sigma, \tau, \bk}} u_{-\sigma, \tau, \bk}^*
\label{C2TFB}
\eeq
The phase factors $\eta_{\sigma,\tau,\bk}$ are not completely free since we want to preserve the relation $C_2 \T R C_2 \T = -R$ satisfied by the microscopic operators which implies that $\eta_{-\sigma,-\tau,\bk} = \eta_{\sigma,\tau,\bk} + \pi$. Otherwise, we have the freedom to choose the relative phase between the wavefunctions in sublattices A and B such that $\eta_{A,K,\bk} = \eta_{B,K,\bk} = 0$ which implies $\eta_{A,K',\bk} = \eta_{B,K',\bk} = \pi$. This means that $C_2 \T$ acts as $\sigma_x \tau_z \K$ in the flat band basis. Note that the action of any of the remaining symmetries, e.g. $\T$ or $C_2$, on the flat band will contain extra phase factors that cannot be removed. In particular, it can be shown that the band topology enforces a non-trivial phase factors for $C_2$ symmetry \cite{KIVCpaper}. %\EK{Don't know if this last comment is necessary}

Using Eq.~\ref{RFB}, we can relate the diagonal entries of the form factor between $\sigma, \tau$ and $-\sigma,-\tau$
\beq
\langle u_{\sigma, \tau, \bk}| u_{\sigma, \tau, \bk + \bq} \rangle = \langle u_{\sigma, \tau, \bk}| R^2 | u_{\sigma, \tau, \bk + \bq} \rangle = \langle u_{-\sigma, -\tau, \bk}| u_{-\sigma, -\tau, \bk + \bq} \rangle, 
\eeq
This can be used to relate the form factors in opposite valleys at the same momentum. In addition, we can use Eq.~\ref{C2TFB} to related the form factors within the same valley for the two sublattices leading to
\beq
\langle u_{\sigma, \tau, \bk}| u_{\sigma, \tau, \bk + \bq} \rangle = \langle u_{\sigma, \tau, \bk}| (C_2 \T)^2| u_{\sigma, \tau, \bk + \bq} \rangle = \langle u_{-\sigma, \tau, \bk}| u_{-\sigma, \tau, \bk + \bq} \rangle^* 
\eeq
Thus, the diagonal entries of the form factor are related by complex conjugation upon exchanging sublattice or valley but remain the same under exchanging both. This leads to the following simple form of the form factor matrix
\beq
\Lambda_\bq(\bk) = F_\bq(\bk) e^{i \Phi_\bq(\bk) \sigma_z \tau_z}
\label{Lambdaqk}
\eeq
where $F_\bq(\bk)$ and $\Phi_\bq(\bk)$ are scalars. Note that the form factor of a given flat band only depends on the combination $\sigma_z \tau_z$ which corresponds to the Chern number. This flips under exchanging sublattice (under $C_2 \T$ symmetry), or valley (under $\T$) but remains invariant under flipping both (under $C_2$ symmetry). %\EK{Maybe I'll add a figure here}

\subsection{Symmetries and the Kinetic Term} 
\label{Sec:oneparticleSymmetries}
 Different symmetries restrict the form of the single particle term $h(\bk)$. Our main assumption is that, although its value may be renormalized by interactions, it has the same symmetries as the projected chiral BM Hamiltonian (\ref{Hint}). First, we can again use spin and valley symmetries to deduce $h(\bk)$ is proportional to $s_0$ and diagonal in valley index. For a single valley, $h(\bk)$ is stricly off-diagonal in sublattice index since:
\beq
\langle u_{\sigma, \tau, \bk}| H_{\rm BM}(\bk) | u_{\sigma, \tau, \bk} \rangle = -\langle u_{\sigma, \tau, \bk}| \sigma_z H_{\rm BM}(\bk) \sigma_z | u_{\sigma, \tau, \bk} \rangle = -\langle u_{\sigma, \tau, \bk}| H_{\rm BM}(\bk) | u_{\sigma, \tau, \bk} \rangle = 0
\eeq
This means that the matrix $h(\bk)$ can only contain the terms $\sigma_x, \sigma_y, \sigma_x \tau_z$, and $\sigma_y \tau_z$. We can use the $R$ symmetry to restrict this form further using  
\beq
\langle u_{\sigma, \tau, \bk}| H_{\rm BM}(\bk) | u_{\sigma', \tau', \bk} \rangle = \langle u_{\sigma, \tau, \bk}| R H_{\rm BM}(\bk) R | u_{\sigma', \tau', \bk} \rangle = \langle u_{-\sigma, -\tau, \bk}| H_{\rm BM}(\bk) | u_{-\sigma', -\tau', \bk} \rangle
\eeq
which implies that $h(\bk)$ commutes with $\sigma_x \tau_x$ excluding the terms $\sigma_y$ and $\sigma_x \tau_z$. This leads to the form
\beq
h(\bk) = h_x(\bk) \sigma_x + h_y(\bk) \sigma_y \tau_z
\eeq

\subsection{Deviations from the chiral limit}
\label{sec:DeviationFromChiralLimit}
Let us now consider what happens when we deviate from the chiral limit $w_0 = 0$. In this case, chiral symmetry is broken which means that the unitary symmetry $R$ is also broken. As a result, we do not expect the form factors to be diagonal in the sublattice index since the wavefunctions are no longer completely sublattice polarized. We can still use valley and spin symmetries to show that the form factor is diagonal in valley and proportional to the identity in spin space.

To restrict the form factor further, we note that the particle-hole symmetry $\P$ remains a symmetry to a good approximation provided we assume the angle is small \footnote{it is only a symmetry if we ignore the angular rotation in the Pauli matrices in Eq.~\ref{HUUDD} which is equivalent to the replacemntsuch that $\sigma_{\pm \theta/2} \mapsto \sigma$}. We note that the product of $\P$ and $\T$ anticommutes with the single particle Hamiltonian which means that the action of $\P \T = \mu_y \tau_x \sigma_x$ maps positive energy states to negative energy states at the same momentum $\bk$. As a result, it leave the space of flat bands invariant and we can write
\beq
\P \T u_{\sigma,\tau,\bk} = e^{i \varphi_{\sigma, \tau, \bk}} u_{-\sigma, -\tau, \bk}
\eeq
To be compatible with the gauge choice in  Appendix \ref{Sec:twoparticleSymmetries}, we choose the gauge such that $\P \T = i \sigma_z R = \sigma_y \tau_x$ which implies that the form factor satisfies $\sigma_y \tau_x \Lambda_\bq(\bk) \sigma_y \tau_x = \Lambda_\bq(\bk)$. This restricts the form factor to the terms $\tau_0 \sigma_{0,y}$ and $\tau_z \sigma_{x,z}$. In addition, we can use $C_2 \T$ (cf.~Eq.~\ref{C2TFB}) to get
\beq
\langle u_{\sigma, \tau, \bk}| u_{\sigma', \tau, \bk+\bq} \rangle = \langle u_{\sigma, \tau, \bk}| (C_2 \T)^2| u_{\sigma', \tau, \bk+\bq} \rangle = \langle u_{-\sigma, \tau, \bk}| u_{-\sigma', \tau, \bk+\bq} \rangle^*
\eeq
which yields $\Lambda_\bq(\bk) = \sigma_x \Lambda_\bq(\bk)^* \sigma_x$ which implies that the coefficients of $\tau_0 \sigma_0$, $\tau_0 \sigma_x$ and $\tau_z \sigma_y$ are real while the coefficient of $\tau_z \sigma_z$ is imaginary leading to the form
\beq
\Lambda_\bq(\bk) = F_\bq(\bk) e^{i \Phi_\bq(\bk) \sigma_z \tau_z} + \sigma_x \tau_z \tilde F_\bq(\bk) e^{i \tilde \Phi_\bq(\bk) \sigma_z \tau_z}
\eeq
Compared to Eq.~\ref{Lambdaqk}, there is an extra sublattice off-diagonal term which is only non-vanishing away from the chiral limit.

\section{Perturbative correction to energy due to single-particle disperison $h$}
\label{app:Tunneling}
To compute the perturbative correction to energy due to single-particle dispersion $h$, we write $\Delta E = J (\bn_+ \cdot \bn_- - 1)$ and take $\bn_+ = -\bn_- = \hat z$. We see that the contribution from the tunneling $\Delta E$ is equal to twice the contribution from tunneling connecting a single fully filled Chern band to a single fully empty band in the opposite Chern sector. This means we can restrict ourselves to a pair of tunnel-coupled opposite Chern bands and just multiply the result by 2 \footnote{the factor of 2 cancels out due to our definition of $J$ with $\Delta E = - 2 J$ such that the absolute value of the result is precisely J}. For definiteness, we take the positive Chern band to be fully filled and the negative Chern band to be fully empty such that the total number of electrons in the $\pm$ band is $N_+ = N$ and $N_- = 0$. The  single particle ``tunneling" term has the form
\beq
\hat h = \sum_\bk c_\bk^\dagger h(\bk) c_\bk = \sum_\bk c_{+,\bk}^\dagger [h_x(\bk) - i h_y(\bk)] c_{-,\bk} + \text{h.c.}
\eeq
where $c_\bk = (c_{+,\bk}, c_{-,\bk})$  is a 2-component vector with annihilation operators in the $\pm$ Chern band. $\hat h$ does not conserve the particle number in each Chern sector separately. Acting with in on a state with $(N_+, N_-) = (N, 0)$ yields a state with $(N_+, N_-) = (N-1, 1)$ with an electron hole pair at some momentum $\bk$ which can be written as
\beq
|\Psi_\bk \rangle = c^\dagger_{-,\bk} c_{+,\bk} |\Psi_0 \rangle
\eeq
The interacting part of the Hamiltonian conserves momentum and Chern number so it maps the state $|\Psi_\bk \rangle$ to a state $|\Psi_{\bk'} \rangle$ and it determines the energy spectrum of the particle-hole excitations. As a result, the second order correction to the energy can be written as
\beq
\Delta E = - \frac{1}{N} \sum_{\bk, \bk'} [h_x(\bk) + i h_y(\bk)] [\H_{\rm eh}]_{\bk,\bk'}^{-1} [h_x(\bk) - i h_y(\bk)], 
\eeq
with $\H_{\rm eh}$ given by \cite{KIVCpaper}
\beq
[\H_{\rm eh}]_{\bk,\bk'} = \frac{1}{2A} \sum_\bq V_\bq \langle \Psi_\bk| \delta \rho_\bq \delta \rho_{-\bq} | \Psi_{\bk'} \rangle = \frac{1}{A} \sum_\bq V_\bq F_\bq(\bk)^2 [\delta_{\bk,\bk'} - \delta_{\bk',[\bk+\bq]} e^{2 i \Phi_\bq(\bk)}]
\eeq
where $[\bq]$ denotes the part of $\bq$ within the first Moir\'e Brillouin zone.

\bibliographystyle{ieeetr}
\bibliography{refs}

\end{document}